\documentclass[preprint,floatfix,aps,prd,showpacs,amsmath,amssymb,amsfonts,superscriptaddress,nofootinbib]{revtex4-1}
\usepackage{graphicx,mathrsfs,float}
\usepackage{color}
\usepackage{enumitem}
\usepackage{dsfont}
\usepackage[normalem]{ulem}
\usepackage{hyperref}
\usepackage{mathtools}
\usepackage{comment}
\usepackage{tensor}
\usepackage[dvipsnames]{xcolor}
\usepackage{soul}
\usepackage{mathtools,braket}
\usepackage[utf8]{inputenc}
\usepackage{multirow} 
\usepackage{subfigure}
\usepackage{varioref}
\usepackage{adjustbox}
\usepackage{rotating}
\usepackage{tensor}
\usepackage{multirow}
\usepackage[active]{srcltx}

\expandafter\ifx\csname package@font\endcsname\relax\else
\expandafter\expandafter
\expandafter\usepackage
\expandafter\expandafter
\expandafter{\csname package@font\endcsname}%
\fi
\allowdisplaybreaks
\hypersetup{
	colorlinks=true,       
	linkcolor=blue,          
	citecolor=blue,        
}

\usepackage{array}
\newcolumntype{P}[1]{>{\centering\arraybackslash}p{#1}}

\usepackage[section]{placeins}

\usepackage{tikz}

\usetikzlibrary{positioning}

\usepackage{fancybox}

\newcommand{\bea}{\begin{aligned}}
\newcommand{\eea}{\end{aligned}}
\def\bea{\begin{eqnarray}}
\def\eea{\end{eqnarray}}
\def\beq{\begin{equation}}
\def\eeq{\end{equation}}
\def\bse{\begin{subequations}}
\def\ese{\end{subequations}}

\newcommand{\D}{\mathscr{D}_x}
\begin{document}
\title{Resolution of Infrared Entanglement Divergences via the Extended Uncertainty Principle}
\author{Subhra Mondal}
\email{subhra.mondal@iitb.ac.in}
%
\author{S. Shankaranarayanan}
\email{shanki@iitb.ac.in}
\affiliation{Department of Physics,  Indian Institute of Technology Bombay, Mumbai 400076, India}


\begin{abstract}
The entanglement entropy of quantum systems typically exhibits both ultraviolet and infrared (IR) divergences. In the low-frequency limit, the IR divergence is intimately tied to the unbounded spatial delocalization of zero-modes, a pathological feature common to both coupled harmonic oscillators and massless scalar fields. In this work, we demonstrate that this infinite growth is naturally resolved by invoking the Extended Uncertainty Principle (EUP), which introduces large-length-scale geometric corrections to the canonical commutation relations. By exactly solving the simple harmonic oscillator under the EUP framework, we establish the existence of an intrinsic geometric confinement that enforces a strict upper bound on the position variance, limits spatial delocalization, and introduces an intrinsic localization length scale related to the background Ricci scalar. We extend this regularizing mechanism to many-body systems by evaluating the entanglement entropy and entanglement spectrum of a one-dimensional harmonic chain and a massless scalar field. We show that the EUP-induced spatial bounds prevent the accumulation of low-lying long-wavelength modes, keeping the entanglement spectrum discrete and evenly gapped even in the strictly massless limit. This non-vanishing modular gap effectively caps the local entanglement temperature of the vacuum. Consequently, the entanglement entropy saturates to a finite value, providing a robust, geometric resolution to the zero-mode IR divergence problem in quantum field theory.
\end{abstract}
 
\maketitle

\section{Introduction}

Entanglement entropy serves as a fundamental measure of quantum correlations in many-body systems and quantum field theories~\cite{Bombelli:1986rw,Srednicki:1993im, 2009Latorre.RieraJ.Phys.A,2009Horodecki.etalRev.Mod.Phys.,Das:2008sy,2010-Eisert.etal-Rev.Mod.Phys.,2010VieiraJournalofPhysicsConferenceSeries,2011-Solodukhin-LivingRev.Rel.,2012Sachdev,2016SavaryReportsonProgressinPhysics}. For lattice-regularized quantum fields of mass $m_f$, the theoretical calculation of entanglement entropy naturally encounters divergent terms that require regulation via a short-distance ultraviolet (UV) cutoff (such as the lattice spacing $d$) and a large-distance infrared (IR) cutoff (the finite system size $L=Nd$). While the leading-order area-law divergence is UV-sensitive and accurately reproduces field theory results, the subleading IR-sensitive terms remain highly non-trivial, particularly in the low-energy limit where the field approaches a massless state~\cite{mallayadiv,chandran_divergence_2019,chandran_one--one_2020}.

Fundamentally, the divergence of entanglement entropy in this IR regime is a purely spatial phenomenon driven by unbounded position variance. As the effective mass of a field approaches zero ($m_f \to 0$), or equivalently, as the frequency of a coupled harmonic oscillator mode approaches zero ($\omega \to 0$), the restoring force vanishes. Consequently, the ground state wavefunction undergoes severe spatial delocalization ($\braket{x^2} \to \infty$). This unbounded spread allows for the proliferation of zero-modes, generating infinite-range quantum correlations across the system partitions that ultimately drive the entropy to diverge. 

In the literature, the exact nature of this IR divergence depends heavily on the system's boundary conditions and artificial IR regulators~\cite{chandran_one--one_2020,Jain:2021ppx}. In the case of conformal field theories in $(1+1)$-dimensions, the subleading term relative to the UV-terms is a non-universal constant~\cite{2005Casini.HuertaJ.Stat.Mech.,2009Casini.HuertaJ.Phys.A,mallayadiv,2009-Calabrese.Cardy-JournalofPhysicsAMathematicalandTheoretical,2016Bianchini.CastroAlvaredoNucl.Phys.B}, whereas in a generic quantum field theory, a $\log$ or $\log-\log$ type divergent behaviour can manifest in the presence of a large number of near zero-modes~\cite{2005Casini.HuertaJ.Stat.Mech.,2009Casini.HuertaJ.Phys.A,mallayadiv,chandran_divergence_2019,chandran_one--one_2020}. Previous works have shown that these IR-sensitive terms capture signatures of a quantum crossover at around $m_f L \sim \mathscr{O}(1)$~\cite{Jain:2021ppx}, indicating an intense competition between the intrinsic spatial delocalization scale of the field ($1/m_f$) and the artificial bounding box of the system ($L$). When $m_f L \ll 1$, the overlap function and entanglement entropy become highly sensitive to the artificial system boundary~\cite{1967AndersonPhys.Rev.Lett.,2006Zanardi.PaunkoviifmmodeacutecelsecfiPhys.Rev.E,2008Zhou.BarjaktarevicJournalofPhysicsAMathematicalandTheoretical,2010VieiraJournalofPhysicsConferenceSeries,2017-Kumar.Shankaranarayanan-SRep,Nenmeli:2022ggg}, highlighting a critical gap: the standard theory lacks an intrinsic physical mechanism to regulate macroscopic spatial fluctuations.

To systematically address this unbounded spatial delocalization without relying on an artificial system size $L$, one must re-examine the foundational limits of quantum mechanics at macroscopic scales. This naturally leads to the Extended Uncertainty Principle (EUP)~\cite{mignemi2010extended,Wagner:2021bqz,Dabrowski:2020ixn,Singh:2021iqa,Gupta:2019cmo,Gattu:2022grl}. Unlike the Generalized Uncertainty Principle (GUP)~\cite{Kempf1995-ka,Nenmeli:2021orl}, which modifies short-distance (UV) physics to introduce a minimum measurable length, the EUP introduces a fundamental modification to the canonical commutation relations at large length scales~\cite{Kempf:1996mv, Park:2007az,Wagner:2021thc,Petruzziello:2021vyf}. Rooted in the background geometry of spacetime---often related to the Ricci scalar---the EUP acts as a geometric metric that imposes a hard mathematical limit on spatial delocalization. By predicting a minimum momentum uncertainty, the EUP enforces an intrinsic maximum correlation length (parameterized by $\gamma^{-1/2}$)~\cite{gattu_extended_2024}. 

In this work, we investigate the resolution of the IR divergence in entanglement entropy through the lens of the EUP. Focusing heavily on the interplay of length scales, we demonstrate that the EUP naturally cures the IR catastrophe by forcing the spatial variance of zero-modes to saturate to a finite geometric bound ($\braket{x^2} \to 1/\gamma$), rather than growing infinitely. 

The paper is organized as follows. In Section~\eqref{section:SHO}, we solve the exact eigenstates of the simple harmonic oscillator (SHO) in the EUP framework, revealing an intrinsic spatial confinement even in the free-particle limit. In Section~\eqref{section:CHO}, we extend this framework to coupled harmonic oscillators. Because the exact EUP ground states exhibit non-Gaussian power-law tails that complicate direct energy evaluation, we employ a Moment-Matched Gaussian Reference State, a rigorous continuous-variable quantum information technique~\cite{Genoni:2008zxa,Adesso:2007jg,Adesso:2014npz,Lami:2018enc}. By strictly enforcing the geometric bounds on spatial correlations dictated by the EUP within the covariance matrix, we leverage the Maximum Entropy Principle to prove that the entanglement entropy saturates to a finite value. Finally, in Section~\eqref{section:Chain} and Section~\eqref{section: fieldTheoryEUP}, we extend this methodology to a 1D interacting chain and its continuum scalar field limit, demonstrating that the geometric regulation provided by the EUP fundamentally shields the field from IR divergences, completely independent of external boundary conditions or finite-size cutoffs. The five  Appendices contain the mathematical details of the key expressions derived in the main text.

\section{Modifications in Momentum and Hamiltonian}
\label{sec:EUP}

As discussed in Ref.~\cite{gattu_extended_2024}, the EUP introduces a fundamental modification to the canonical commutation relation:
\begin{equation}
    [\hat{x}, \hat{p}]=i\hbar(1+\gamma \hat{x}^2) \label{eqn:comm1d}
\end{equation}
Unlike the Generalized Uncertainty Principle (GUP), which dictates very short-distance physics (the UV limit) and predicts a minimal observable length scale, the EUP affects large length scales and predicts a minimum possible momentum. Here, $\gamma$ is a scalar related to the curvature scale of the background metric; in flat spacetime backgrounds, $\gamma=0$. Throughout this calculation, a small finite value of $\gamma$ is assumed. 

Provided that the new commutation relation is identically satisfied, necessary modifications to the standard quantum-mechanical operators must be made. We consider the following scenario: modifying the momentum operator while keeping the position operator unchanged. Crucially, we must also ensure the Hermiticity of this modified operator to preserve a time-reversal-symmetric Hamiltonian.

The following ansatz for the momentum operator is suitable for this case: 
\begin{equation}
\hat{p}= f(\hat{x})\hat{\pi} +g(\hat{x}) \label{eqn:ansatzcomm}
\end{equation}
Here, we introduce $\hat{\pi}$ as the standard momentum operator in flat geometry, satisfying $[\hat{x},\hat{\pi}]=i\hbar$. The modified momentum operator must satisfy Eq.~\eqref{eqn:comm1d}. Note that the operator $g(\hat{x})$ always commutes with $\hat{x}$. Therefore, the commutation relation alone is inconclusive regarding the exact functional forms of $f(\hat{x})$ and $g(\hat{x})$. Imposing the Hermiticity condition uniquely fixes $g(\hat{x})$, yielding the final form of the momentum operator:
\begin{equation}
   \hat{p}= (1+\gamma \hat{x}^2)\hat{\pi} -i\hbar\gamma\hat{x}\label{eqn:mod_p}
\end{equation}
Crucially, the Hermiticity condition requires the boundary surface terms to vanish. The relevant boundary term for the momentum operator is $\left[\psi^*(x)\psi(x) (1+\gamma x^2)\right]_{-\infty}^\infty$. For this to vanish, the probability density must approach zero faster than the geometric factor $(1+\gamma x^2)$ diverges. Provided this condition is met, the momentum operator is strictly Hermitian. As will be shown in the exact solution to the Simple Harmonic Oscillator (SHO) in Section~\eqref{section:SHO}, the slowest decaying part of the asymptotic wavefunction scales as $\psi\sim (1+\gamma x^2)^{-(1+l/2)}$, where $l$ is a real positive number ($l=0$ for a free particle). Hence, this formally verifies that the operator remains Hermitian for both the SHO and free particle systems.

The commutation relation alone does not uniquely fix the representation of the momentum operator. The addition of a real function $h(x)$ corresponds to introducing an additional position-dependent potential term in the Hamiltonian, effectively modifying the system's dynamics. Since our primary goal is to isolate the physical effects arising purely from the modified commutation relation, we restrict ourselves to the minimal representation where no artificial potential-like terms are introduced. This corresponds to choosing $h(x) = \text{constant}$, which can be set to zero without any loss of generality.

\section{Simple Harmonic Oscillator in the EUP Framework}
\label{section:SHO}

One of the simplest yet most elegant systems in quantum mechanics is SHO with mass $M$ and frequency $\omega$. Furthermore, one can construct coupled structures from the primitive SHO to study quantum properties like entanglement. In this section, we solve for the exact energy eigenstates of the SHO within the EUP framework. Because the modified momentum operator is explicitly position-dependent, the analysis is most naturally performed in the position representation. The time-independent Schr\"odinger equation for the SHO, in the EUP framework, takes the following form:
\begin{equation}
\begin{split}
    (1+\gamma x^2) \psi''(x)+ 4\gamma x\psi'(x) +\left[2\gamma + \frac{2M(E-M\omega^2x^2/2)/\hbar^2 -\gamma}{1+ \gamma x^2}\right]\psi(x)=0 \label{eqn:SHO1}
\end{split}
\end{equation}
Using a series of transformations, the above differential equation can be solved exactly by mapping it to the associated Legendre equation~\cite{Olver:2010ouy}. To go about that, first, we introduce the following dimensionless parameters into Eq.~\eqref{eqn:SHO1}: $\gamma = \frac{M\omega}{\hbar}\epsilon$ and $x = \sqrt{\frac{\hbar}{M\omega}}\xi$, which yields:
\begin{equation}
    (1+\epsilon\xi^2)\phi''(\xi) + 4\epsilon\xi \phi'(\xi)+ \left[2\epsilon + \frac{2E/\hbar\omega-\epsilon -\xi^2}{1+\epsilon \xi^2}\right]\phi(\xi) =0
\end{equation}
The second step is to do rewrite $\xi$ in terms of $z$ and $\phi(\xi)$ interms of $W(z)$:
\begin{equation}
 \epsilon\xi^2 = -z^2, \quad \phi(\xi) = (1-z^2)^{-1/2}W(z) 
 \label{def:SHO-subs}
\end{equation}
Substituting these in the above differential equation, leads to: 
\begin{equation}
    (1-z^2)W''(z) -2z W'(z) +\left(\frac{1}{\epsilon^2} - \frac{2E/\hbar\omega\epsilon + 1/\epsilon^2}{1-z^2}\right)W(z)=0\label{eqn:shofinal}
\end{equation}
and is of the form of the associated Legendre equation with the following mapping of the parameters $1/\epsilon^2 = l(l+1)$ and $(2E/\hbar\omega\epsilon) + 1/\epsilon^2 = m^2$. Note that there are no physical conditions imposed yet to restrict the values of $l$ and $m$. In that respect, we assume $l, m \in \mathbb{C}$. 

From the definitions \eqref{def:SHO-subs}, we have: 
$z = \pm i\sqrt{\epsilon}\xi = \pm i\sqrt{\gamma}x$. Thus, we see that the associated Legendre equation possesses singular points at $z = \pm 1$. {However, real positive values of $\gamma$ ensure that $z$ remains strictly imaginary. Thus, no singularities are encountered along the real $x$ line.}

Using the hypergeometric representation of $P_l^m(\pm i\sqrt{\gamma}x)$~\cite{Olver:2010ouy,NIST:DLMF}: 
\begin{equation}
\begin{split}
      P_l^m(\pm i\sqrt\gamma x)
=\left(\frac{1 \pm i\sqrt{\gamma}x}{1 \mp i\sqrt{\gamma}x}\right)^{m/2}
{_2F_1}\!\left(l+ 1,\,-l;\,1 - m;\,\frac{1}{2} \mp \frac{i\sqrt{\gamma}x}{2}
\right) .
\label{eqn: generalsoln}
\end{split}
\end{equation}
one may analytically continue the solution beyond $|z|=1$ if required, with a branch cut chosen along $(1,\infty)$. For details on the analytic continuation for $|z|>1$ see also, Ref.~\cite{lopezNewSeriesExpansions2013}. 

\subsection{Boundary Conditions and Quantization}
Although the Gauss hypergeometric function ${_2F_1}$ contains gamma functions of $l$ and $m$---where integer values of $m>1$ may formally lead to poles---these singularities are simple and can be systematically removed by taking appropriate limits. Consequently, ${_2F_1}$ remains convergent for all real values of $x$ given a specific set of $l$ and $m$. From a physical standpoint, the wavefunction must naturally vanish in the asymptotic limits $x \to \pm\infty$. Imposing this normalizability requirement yields the quantization condition:
$$
m-l\in \mathbb{N}\quad \text{or}\quad m-(l+1)\in \mathbb{Z}^+ .
$$

Accordingly, we choose $m = l + 1 + n$, where $n$ is a non-negative integer. For this specific choice, the wavefunction decays to zero at spatial infinity. This behavior can be demonstrated analytically using the following transformation identity of ${_2F_1}$:
\begin{equation}
    {_2F_1}(a,b;c;z)= (1-z)^{c-a-b}{_2F_1}(c-a, c-b; c;z)
\end{equation}

Substituting our specific parameters ($a = l+1$, $b = -l$, and $c = -l-n$) into the identity yields:
\begin{equation}
    {_2F_1}(l+1,-l; -l-n, z) = \frac{1}{(1-z)^{l+n+1}}{_2F_1}(-2l-1-n,-n; -l-n,z)
\end{equation}
If $n$ is a non-negative integer, the hypergeometric function on the right-hand side reduces to a finite polynomial of degree $n$, which inherently prevents asymptotic divergence. Specifically, the expansion takes the exact form:
\begin{equation}
    {_2F_1}(l+1,-l; -l-n, z) = \frac{1}{(1-z)^{l+n+1}}\sum_{k=0}^n \frac{(-n)_k(-2l-n-1)_k}{(-l-n)_k\ k!} z^k
\end{equation}
The parameter $l$ is strictly determined by the physical constants $\gamma$ and $\omega$. To avoid multi-valued functions and poles within the physical domain, we enforce these mathematical constraints. Incorporating these assumptions, the final form of the unnormalized energy eigenfunctions is given by:
\begin{equation}
    \psi_n(x) = \frac{N_n}{(1+\gamma x^2)^{1+(l+n)/2}} \sum_{k=0}^n \frac{(-n)_k \, (-2l-n-1)_k}{(-l-n)_k \, k! \, 2^k} (1 \mp i\sqrt{\gamma} x)^k \label{eqn:eigstates}
\end{equation}
Here, $N_n$ is the normalization constant (see Appendix~\eqref{appendix:normalization} for its closed-form expression), and $l$ satisfies the relation $l(l+1) = \left(\frac{M\omega}{\hbar\gamma}\right)^2$. The fully normalized ground state is given by:
\begin{equation}
    \psi_0(x)=\left(\frac{\gamma}{\pi}\right)^{1/4}\left(\frac{\Gamma(l+2)}{\Gamma(l+3/2)}\right)^{1/2} \frac{1}{(1+\gamma x^2)^{1+l/2}} 
\end{equation}
Unlike the rapid Gaussian decay characteristic of the standard SHO, the EUP-modified ground state exhibits a heavy power-law tail. This modified asymptotic behavior ensures normalizability even in the complete absence of an external confining potential. Consequently, in the strict free-particle limit, the ground-state wavefunction naturally converges to a Lorentzian distribution. 

\begin{figure}[!htb]
    \centering
    \includegraphics[width=0.45\linewidth]{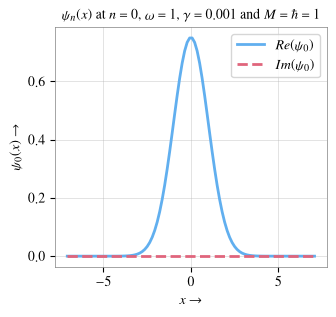}
    \includegraphics[width=0.45\linewidth]{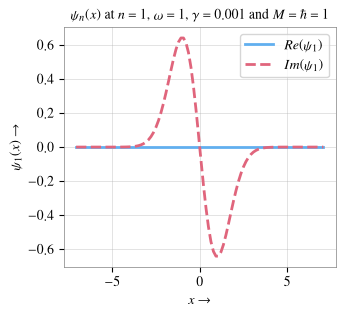}\\
    \includegraphics[width=0.45\linewidth]{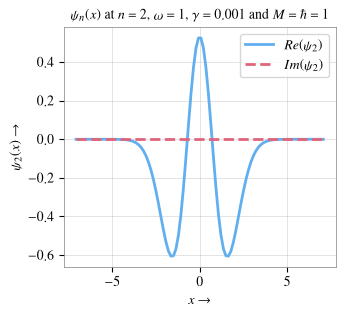}
    \includegraphics[width=0.45\linewidth]{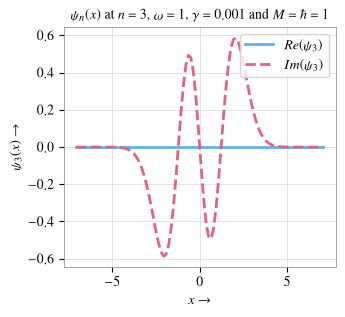}
    \caption{The spatial profiles of the first four discrete eigenstates of the EUP-modified oscillator.}
    \label{fig:Psis}
\end{figure}

An important consequence of the EUP-modified ground state is the emergence of an intrinsic correlation length scale. This is best understood by evaluating the saturation limit of the position variance (derived explicitly in Appendix~\eqref{appendix: vargsw}):
\begin{equation}
   \lim_{\omega\to0} \braket{\hat x^2}=\lim_{l\to0}\frac{1}{\gamma(2l+1)}=\frac{1}{\gamma}
\end{equation}
This mathematically guarantees that the wavefunction cannot spread arbitrarily over infinite space. Instead, it is geometrically restricted by an effective confinement scale of order $\gamma^{-1/2}$, beyond which the probability density is strongly suppressed. 

The first few eigenstates are depicted in Fig.~\eqref{fig:Psis}. From Eq.~\eqref{eqn:eigstates} we see that there are two choices of $z = \pm i \sqrt{\gamma} x$,
indicating that the states are doubly degenerate. However,  expanding and simplifying the summation reveals that the physical solutions are purely real or purely imaginary for even and odd $n$, respectively. Consequently, $\psi_n^+$ and $\psi_n^-$ are not linearly independent, confirming that the bound states are \emph{fundamentally non-degenerate}.

Based on the structure of the wavefunction, we can identify two distinct physical regimes:
\begin{description}
    \item[\textbf{The HUP Limit ($M\omega/\hbar\gamma \gg 1$)}] In this regime, the geometric deformation in the kinetic sector is negligible compared to the external potential energy. Consequently, standard quantum mechanical behavior dictates the system's dynamics, effectively masking the EUP corrections.
    \item[\textbf{The EUP Limit ($M\omega/\hbar\gamma \ll 1$)}] In this low-frequency regime, the EUP effects decisively take over. In the strict free-particle limit, the Lorentzian structural tail enforces an intrinsic localization that is completely absent in standard unbounded continuous frameworks.
\end{description}

\subsection{Energy Spectrum of SHO }

Equation \eqref{eqn:shofinal} yields the following general expression for the energy eigenvalues:
\begin{equation}
    E=\frac{\hbar\omega\epsilon}{2}\left[m^2-l(l+1)\right]
\end{equation}
Imposing the required quantization condition $m = l + n + 1$ results in a discrete, exact energy spectrum:
\begin{equation}
    E_n=\hbar\omega l\epsilon \left(n+\frac{1}{2}\right)+\frac{\hbar \omega\epsilon(n+1)^2}{2}
\end{equation}

This explicit $n^2$ dependence is characteristic of a particle-in-a-box spectrum, formally indicating the presence of an effective geometric bounding length scale over which the particle is confined. This physical insight motivates a deeper investigation into the specific limits of the theory. To systematically understand the interplay between the external harmonic potential and the intrinsic geometric deformation, we analyze the exact energy spectrum and the corresponding wavefunctions across three distinct physical regimes.

\subsubsection{The Standard Flat-Space Limit ($\gamma \to 0$)}
In the strict standard limit where the background curvature vanishes ($\gamma \to 0$), the geometric deformation is entirely removed. Mathematically, this corresponds to $l\epsilon \to 1$ and $\epsilon \to 0$. In this limit, the quadratic correction term in the energy spectrum vanishes identically, and the conventional, equally-spaced SHO spectrum is immediately recovered:
\begin{equation}
    E_n = \hbar\omega\left(n+\frac{1}{2}\right) .
\end{equation}

We can rigorously verify this reduction by analyzing the exact spatial ground state. Taking the $\gamma \to 0$ limit of the unnormalized EUP wavefunction yields:
\begin{equation}
    \lim_{\gamma \to 0} \psi_0(x) = \lim_{\gamma\to 0}\frac{N}{1+\gamma x^2}(1+\gamma x^2)^{-M\omega/2\hbar\gamma} .
\end{equation}
Introducing the parameter $s = 1/\gamma$, this limit can be rewritten identically as the definition of an exponential decay:
\begin{equation}
\lim_{s \to \infty}\left(1+\frac{x^2}{s}\right)^{-\frac{M\omega}{2\hbar}s}
    =\exp\left(-\frac{M\omega}{2\hbar}x^2 \right) .
\end{equation}
This expression coincides exactly with the standard Gaussian ground-state wavefunction of the continuous SHO, confirming that standard quantum mechanics is smoothly recovered when the geometric cutoff is removed.

\subsubsection{HUP Regime: The Weak-Deformation Limit ($M\omega/\hbar\gamma \gg 1$)}
In this regime, the system is dominated by the external harmonic potential, but the macroscopic spacetime geometry introduces a small, non-vanishing perturbative correction ($\gamma > 0$). Substituting $l \sim \epsilon^{-1} = M\omega/\hbar\gamma$ into the exact spectrum yields the standard SHO spectrum alongside a leading-order quadratic geometric correction:
\begin{equation}
    E_n = \hbar\omega\left(n+\frac{1}{2}\right) + \frac{\hbar^2\gamma}{2M}(n+1)^2 .
    \label{eqn:energy2}
\end{equation}
This specific correction formally breaks the equal energy spacing of the harmonic oscillator. Notably, this exact leading-order shift can also be independently derived via a semi-classical approximation followed by the quantization of the action, as demonstrated in Ref.~\cite{Hamil:2023zsf}. The explicit $(n+1)^2$ dependence indicates that higher energy states \emph{feel} the geometric bounding more strongly than the lower energy states.

\subsection{Free-Particle Limit in the EUP}
\label{subsec: free-particle-eup}

In addition to the simple harmonic oscillator, the free particle in the EUP framework warrants an explicit analysis rather than being treated merely as a limiting case. The results established here will be key for constructing the scalar field theory in the EUP framework in Sec.~\eqref{section: fieldTheoryEUP}. 

Setting $\omega = 0$ in the EUP Schr\"odinger equation yields the spatial wave functions in the following form:
\begin{equation}
    \varphi(x) = \frac{N}{\sqrt{1 + \gamma x^2}} \exp\left[ \pm i m \arctan\left(\sqrt{\gamma} x\right) \right] ,
\end{equation}
where $N = \gamma^{1/4} / \sqrt{\pi}$ is the continuum normalization constant in the infinite-domain limit $R \to \infty$ (see Appendix~\eqref{appendix: free-particle-details} for explicit finite-box normalization derivations).

To construct solutions under specific boundary conditions, we take linear combinations of these modes into parity-definite states:
\begin{equation}
    \psi(x) = \frac{\gamma^{1/4}}{\sqrt{\pi(1+\gamma x^2)}} \left\{ A \exp\left[i m \arctan\left(\sqrt{\gamma} x\right)\right] + B \exp\left[-i m \arctan\left(\sqrt{\gamma} x\right)\right] \right\} .
    \label{eqn: freeparticleGnrlEigenstatesMain}
\end{equation}

\subsubsection{Dirichlet Boundary Conditions}
Imposing Dirichlet boundary conditions $\psi(\pm R) = 0$ over a domain $x \in [-R, R]$ discretizes the parameter $m$. In the continuum limit ($R \to \infty$), the wave functions simplify to the compact parity-separated form:
\begin{equation}
    \psi_n(x) = \begin{cases}
         \dfrac{\gamma^{1/4}\sqrt{2}}{\sqrt{\pi(1+\gamma x^2)}} \cos\left[n \arctan\left(\sqrt{\gamma} x\right)\right] , & \text{if } n \text{ is odd,} \\[1.2em]
         \dfrac{i\gamma^{1/4}\sqrt{2}}{\sqrt{\pi(1+\gamma x^2)}} \sin\left[n \arctan\left(\sqrt{\gamma} x\right)\right] , & \text{if } n \text{ is even,}
    \end{cases}
\end{equation}
where $n \in \mathbb{N}$ indexes the discrete spatial modes.

\subsubsection{Neumann Boundary Conditions}
Imposing Neumann boundary conditions $\psi'(\pm R) = 0$ is of primary interest for the scalar field theory quantization. As detailed in Appendix~\eqref{appendix: free-particle-details}, the eigenvalue condition yields $m \to n \in \mathbb{N}$ as $R \to \infty$, giving the normalized eigenfunctions:
\begin{equation}
    \psi_n(x) = \begin{cases}
         \dfrac{\gamma^{1/4}\sqrt{2}}{\sqrt{\pi(1+\gamma x^2)}} \cos\left[n \arctan\left(\sqrt{\gamma} x\right)\right] , & \text{if } n \text{ is odd,} \\[1.2em]
         \dfrac{i\gamma^{1/4}\sqrt{2}}{\sqrt{\pi(1+\gamma x^2)}} \sin\left[n \arctan\left(\sqrt{\gamma} x\right)\right] , & \text{if } n \text{ is even.}
    \end{cases}
\end{equation}
Unlike the standard flat-space continuum, the EUP framework naturally dictates a discrete mode spectrum even in the absence of an external potential, serving as an effective geometric regulator for the field theory.

\subsubsection{Energy Spectrum and Geometric Localization}
\label{subsection: freeparticle-limit}

Setting $\omega = 0$ in the EUP wave equation removes the external harmonic potential. However, the exact spatial wave functions and energy spectrum exhibit several distinct features that differ qualitatively from standard non-relativistic quantum mechanics.

As detailed above (and derived explicitly in Appendix~\eqref{appendix: free-particle-details}), imposing either Dirichlet or Neumann boundary conditions on a finite domain $x \in [-R, R]$ and taking the continuum limit $R \to \infty$ yields identical normalized eigenfunctions:
\begin{equation}
    \psi_n(x) = \begin{cases}
         \dfrac{\gamma^{1/4}\sqrt{2}}{\sqrt{\pi(1+\gamma x^2)}} \cos\left[n \arctan\left(\sqrt{\gamma} x\right)\right] , & \text{if } n \text{ is odd,} \\[1.2em]
         \dfrac{i\gamma^{1/4}\sqrt{2}}{\sqrt{\pi(1+\gamma x^2)}} \sin\left[n \arctan\left(\sqrt{\gamma} x\right)\right] , & \text{if } n \text{ is even.}
    \end{cases}
\end{equation}

Because both sets of boundary conditions agree identically in the $R \to \infty$ limit, the resulting energy spectrum is uniquely determined without ambiguity. Setting $\omega = 0$ leads to the geometric term:
\begin{equation}
    E_n = \frac{\hbar^2 \gamma}{2M}n^2 \, , \qquad n \in \mathbb{N} \, .
\end{equation}

\begin{figure}[htb]
    \centering
    \includegraphics[width=0.8\linewidth]{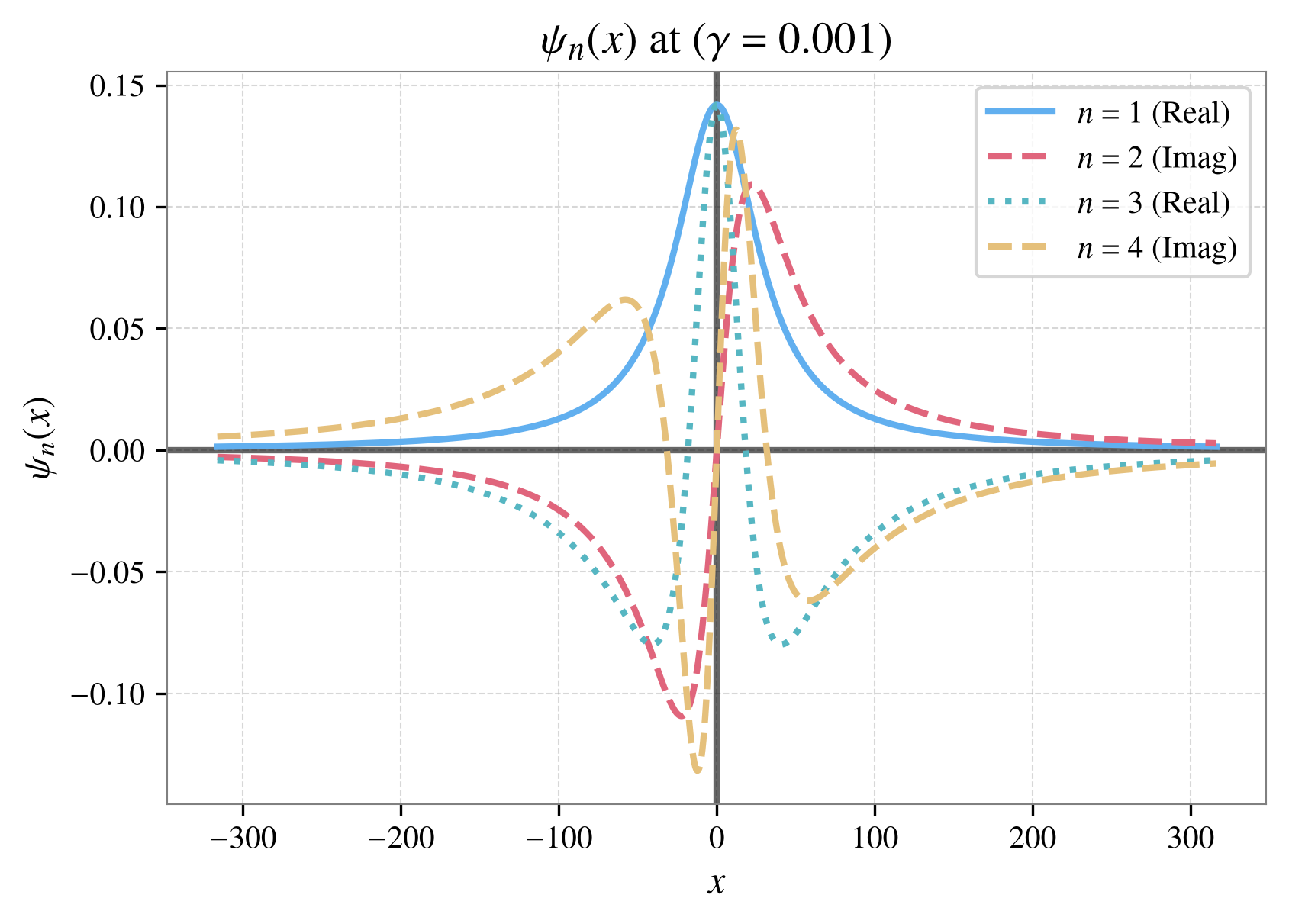}
    \caption{Spatial wave functions of the EUP-deformed free particle.}
    \label{fig:placeholder0}
\end{figure}

This quadratic dependence on $n$ mirrors the energy spectrum of a particle trapped in a box with an effective width on the order of $L \sim 1/\sqrt{\gamma}$. The crucial physical distinction, however, lies in the nature of the corresponding quantum states. In contrast to a standard continuous free particle, these states cannot be characterized by a single, sharply defined wavelength. For example, near the origin where the deformation is weak ($\sqrt{\gamma} x \to 0$), the asymptotic behavior $\arctan(\sqrt{\gamma}x) \approx \sqrt{\gamma} x$ gives rise to an effective local wavenumber $k_n = n\sqrt{\gamma}$. However, as the coordinate distance from the origin increases, this effective wavenumber flattens out. Furthermore, the envelope function $(1+\gamma x^2)^{-1/2}$ dynamically suppresses the amplitude of the wave function far from the origin. These features are illustrated in Fig.~\eqref{fig:placeholder0}.

This lack of monochromaticity is fundamentally consistent with the EUP, as the modified algebra mandates a minimum momentum uncertainty. Consequently, a free particle characterized by an infinite spatial spread and a definite momentum simply does not exist within the EUP framework. Instead, the particle is \emph{intrinsically localized}, behaving as if it were confined within an effective spatial box, even without a background potential.

Importantly, there are no hard physical barrier walls present in the system. Since, the momentum deformation is explicitly coordinate-dependent, long-wavelength spatial fluctuations are dynamically suppressed. This renders spatial infinity inaccessible to the quantum state, allowing the intrinsic geometry to act as a \emph{soft effective} bounding barrier.

\begin{figure}[htb]
    \centering
    \includegraphics[width=0.5\linewidth]{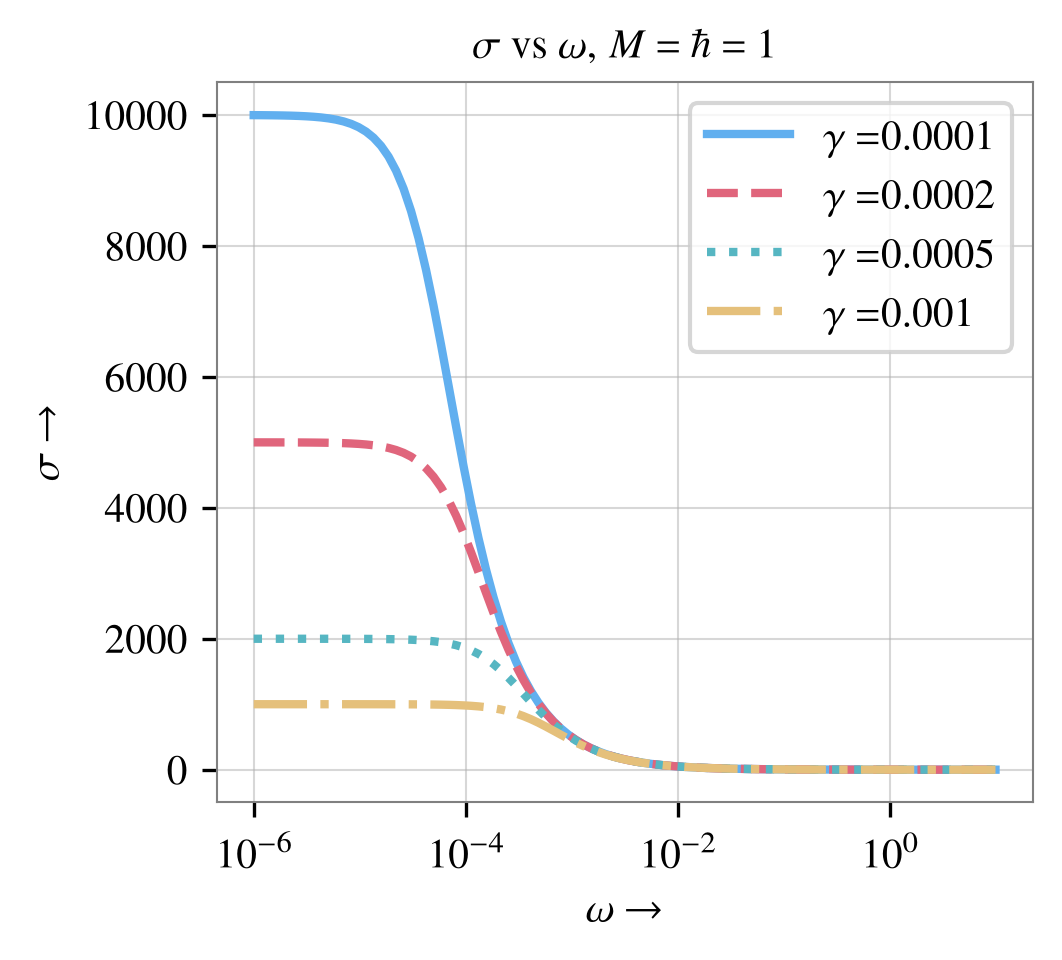}
    \caption{The spatial variance of the ground state as a function of the oscillator frequency $\omega$.}
    \label{fig:varvsomega}
\end{figure}

This intrinsic geometric localization motivates a rigorous examination of the spatial variance of the wave functions. The position variance of an arbitrary free-particle state evaluates to a constant, $\braket{\hat{x}^2} - \braket{\hat{x}}^2 = 1/\gamma$. Alternatively, this free-particle variance can be recovered from the simple harmonic oscillator as a limiting case $\omega \to 0$. Figure~\eqref{fig:varvsomega} illustrates the position variance of the harmonic oscillator as a function of the trapping frequency $\omega$:
\begin{equation}
    \sigma_x^2 = \frac{1}{\gamma(2l+1)} \, , \qquad \text{where} \quad l(l+1) = \left(\frac{M\omega}{\hbar\gamma}\right)^2 .
\end{equation}

The key steps in this derivation are presented in Appendix~\eqref{appendix: vargsw}. This variance saturation is entirely absent in standard quantum mechanics and yields highly nontrivial, regularizing consequences when extended to coupled field-theoretic degrees of freedom. For example, the zero modes in the standard Heisenberg Uncertainty Principle (HUP) framework are completely delocalized, giving rise to IR divergences in the entanglement entropy. The geometric localization induced by the EUP suggests that the entanglement entropy should be capped at a maximum value corresponding to the saturated variance limit $1/\gamma$. We demonstrate this explicitly in the subsequent sections.

This significant departure from continuous translation symmetry stems directly from the modified uncertainty relation itself. As spatial fluctuations grow and approach the characteristic geometric length scale $L \sim \gamma^{-1/2}$, the momentum uncertainty begins to grow rather than shrink. From an algebraic perspective, the explicit position-dependence of the modified momentum operator breaks continuous spatial translation symmetry. As a result, an intrinsic, finite spatial spread is naturally enforced across the system, definitively resolving infinite-volume IR divergences even in the complete absence of an external confining potential.

\section{Coupled Harmonic Oscillators}
\label{section:CHO}

In the standard treatment of coupled harmonic oscillators, the divergence of entanglement entropy in the low-frequency (IR) limit is closely tied to the behavior of the center-of-mass mode~\cite{mallayadiv,chandran_divergence_2019}. As the effective frequency of this mode approaches zero, the corresponding wavefunction becomes increasingly delocalized, leading to an unbounded growth of position fluctuations and, consequently, a divergence in the entanglement entropy \cite{chandran_divergence_2019,Jain:2021ppx}.

As shown in the previous section, in the EUP-modified scenario, however, the variance of the position operator saturates due to the presence of an intrinsic geometric length scale. This fundamentally prevents the center-of-mass mode from becoming arbitrarily delocalized, thereby naturally regulating the IR behavior of the system and eliminating the associated divergence in the entanglement entropy.

To go about quantifying this, let us define the physical Hamiltonian of two harmonic oscillators (with identical mass $M$ and frequency $\omega$) coupled through a quadratic interaction potential:
\begin{equation}
    \hat H=\frac{\hat p_1^2+\hat p_2^2}{2M} + \frac{1}{2}M \omega^2(\hat x_1^2+\hat x_2^2) +\frac{1}{2}M\omega_{\text{int}}^2(\hat x_1-\hat x_2)^2 \, . 
    \label{eqn: coupledHamiltonian}
\end{equation}
In the standard HUP treatment, one performs a coordinate rotation in configuration space, transforming to normal-mode coordinates in which the Hamiltonian completely decouples. However, the non-linearities introduced by the EUP-modified momentum operator reintroduce cross-couplings among the coordinates upon such a transformation. As a result, the physical Hamiltonian in Eq.~\eqref{eqn: coupledHamiltonian} cannot be decoupled by any linear change of variables.

For mathematical comparison, let us define a standard normal-mode Hamiltonian $H'$:
\begin{equation}
    \hat H'=\frac{\hat p_+^2+\hat p_-^2}{2M} + \frac{1}{2}M \omega_+^2 \hat x_+^2 + \frac{1}{2}M \omega_-^2 \hat x_-^2 \label{eqn: normalModeHamiltonian}
\end{equation}
where $\hat x_\pm = (\hat x_1\pm\hat x_2)/\sqrt{2}$ and the corresponding conjugate momenta $\hat p_\pm$ satisfy the modified algebra $[\hat x_\pm, \hat p_\pm]= i\hbar (1+\gamma\hat x_\pm^2)$. We define the normal-mode frequencies as $\omega_+=\omega$ and $\omega_-=\sqrt{\omega^2+2\omega_{\text{int}}^2}$. While this is the typical transformation used in the HUP framework to exactly decouple $H$, the explicit coordinate-dependent deformation in the EUP framework prevents the kinetic sector from decoupling, even though the potential sector is diagonalized. Therefore, $\hat p_+^2+\hat p_-^2 \neq \hat p_1^2+\hat p_2^2$, and consequently, $\hat H \neq \hat H'$. Up to order $\gamma$, $\hat H'$ yields a residual coupling difference of:
\begin{equation}
    \Delta \hat H = \hat H' - \hat H \sim \frac{\hbar^2\gamma}{2M}\left[ -4x_1x_2 \frac{\partial^2}{\partial x_1\partial x_2}+(x_1^2-x_2^2)\left(\frac{\partial^2}{\partial x_1^2} - \frac{\partial^2}{\partial x_2^2}\right)\right]
\end{equation}
In the standard HUP framework, the coupled Hamiltonian can be exactly diagonalized into normal modes, yielding a joint ground state that is purely Gaussian. However, as demonstrated above, the EUP-modified momentum operators introduce non-linearities that prevent an exact linear decoupling of the kinetic sector. Before addressing this challenge, it is essential to identify the physical and mathematical properties of the exact HUP solution that govern entanglement, while concurrently establishing how the EUP fundamentally deviates from this standard formulation.

The exact ground-state solution of the decoupled Hamiltonian $H'$ in normal-mode coordinates takes the standard Gaussian form:
\begin{equation}
    \Psi_0(x_+, x_-)=\left(\frac{\beta_+\beta_-}{\pi^2} \right)^{1/4}\exp\left(-\frac{\beta_+x_+^2 +\beta_-x_-^2}{2} \right)\label{eqn: gaussianCHO}
\end{equation}
where $\beta_\pm = M\omega_\pm/\hbar$. In standard quantum mechanics, the entanglement divergence arises solely from the unbounded spatial spread of the center-of-mass mode (i.e., $\beta_+ \to 0$ as $\omega \to 0$). As established in Section~\eqref{section:SHO}, the EUP introduces two distinct physical properties: a strict saturation of this spatial spread and a non-Gaussian, power-law asymptotic tail. To construct a plausible and robust approximation of the entanglement entropy in the IR regime, we must rigorously preserve this EUP-induced spatial saturation.

An intuitive approach to evaluating the entanglement entropy would be to use the exact eigenstates of the decoupled Hamiltonian $H'$ (i.e., a product state in the $(+,-)$ basis). However, evaluating the energy variance $\sigma_H^2 = \braket{H^2} - \braket{H}^2$ of the physical Hamiltonian for such a state reveals a critical instability. As derived in Appendix~\eqref{appendix: variance_product_state}, the cross-terms in $H$ introduce higher-order coupled spatial moments. Because the exact EUP wavefunctions exhibit heavy power-law asymptotic decay, these unsuppressed higher-order moments cause the energy variance to become unphysically large ($\sigma_H \gg \braket{H}$). Thus, the exact product state fails to provide a stable baseline.

To circumvent this, we propose a Gaussian reference state modeled exactly after the standard HUP solution (Eq.~\eqref{eqn: gaussianCHO}), but we manually upgrade the spread parameters to their exact EUP-modified counterparts:
\begin{equation}
    \beta_{\pm} \to \beta_{\pm, \text{EUP}} = \frac{\gamma}{2}\left[2l_\pm(\omega_\pm) + 1\right] \, .
\end{equation}

The motivation behind this approach arises purely from the perspective of entanglement. Quantities such as entanglement entropy are governed by the uncertainties arising from quantum correlations between subsystems. These uncertainties are encoded in the position and momentum variances \cite{Bombelli:1986rw,mallayadiv,Jain:2021ppx}. Therefore, the fundamental modifications introduced by the EUP, such as the saturation of these variances, must be incorporated when calculating entanglement. In the standard HUP treatment of coupled harmonic oscillators, the spread parameters are given by $\beta_{\pm}=M\omega_{\pm}/\hbar$, which diverge in the IR limit. However, in the normal-mode coordinates, the system is completely decoupled. Since our primary interest is entanglement, rather than obtaining approximate wavefunctions by minimizing the energy, as is done in the conventional approach, we instead modify the functional form of the spread parameters so that they reproduce the moments governing the informational structure of the system. The corresponding $\beta$ parameters are defined through the position variance. For example, in the HUP case, the position variance satisfies $\sigma_x^2=1/(2\beta)$, implying $\beta=1/(2\sigma_x^2)$. We then replace $\sigma_x$ with the EUP-corrected position variance derived earlier.

To ensure a comprehensive analysis, the energy scale, variance, and residual coupling of this trial wavefunction must be examined in the IR limit. The system's characteristic energy scale is determined by estimating the ground-to-excited state energy gap through Eq.~\eqref{eqn:energy2}. As $\omega \to 0$, the energy gap reduces to approximately $\hbar\omega_{\text{int}} + \hbar^2\gamma/M$. Notably, the residual coupling for the Gaussian reference state approaches zero at $\omega=0$ within the weak-coupling regime ($M\omega_{\text{int}} \ll \hbar \gamma$):
\begin{equation}
\begin{split}
    \braket{\Delta \hat H}&=\frac{\hbar^2\gamma}{M}\left[ \frac{1}{2}-\frac{1}{4}\left(\frac{\beta_-}{\beta_+ } + \frac{\beta_+}{\beta_- }\right)\right]\\
    &=\frac{\hbar^2\gamma}{M}\left[\frac{1}{2}-\frac{1}{4}\left(\sqrt{1+4\left(\frac{\sqrt{2}M\omega_{\text{int}}}{\hbar\gamma}\right)^2}+\frac{1}{\sqrt{1+4\left(\frac{\sqrt{2}M\omega_{\text{int}}}{\hbar\gamma}\right)^2} }\right)\right]\approx 0
\end{split}
\end{equation}

\begin{figure}[H]
    \centering
    \includegraphics[width=0.45\linewidth]{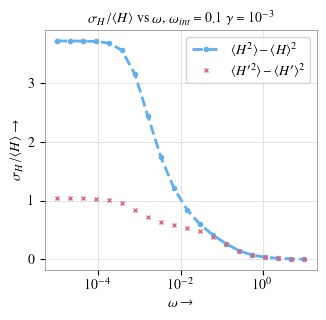}
    \includegraphics[width=0.45\linewidth]{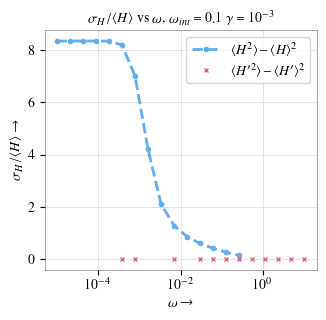}
    \caption{The energy variance of the Gaussian reference state (left) compared to the exact normal-mode Product state (right).}
    \label{fig: H_variance}
\end{figure}

Although we cannot choose arbitrarily small values for the ratio $\omega_{\text{int}}/\gamma$ (since the interaction $\omega_{\text{int}}$ is the fundamental source of entanglement), the emergence of this minimal coupling in the IR regime strongly reinforces the validity of our Gaussian approximation. Furthermore, the Gaussian reference state natively converges to the exact eigenstate of both $\hat H$ and $\hat H'$ in the HUP limit. However, due to the omission of the EUP's heavy tails, the relative energy variance $\sigma_H/\braket{\hat H}$ for the Gaussian state consistently remains of order $\mathcal{O}(1)$ across the spectrum of interaction strengths. Conversely, the fluctuations corresponding to the product state become anomalously large, echoing our prior analytical assessment, as clearly depicted in Fig.~\eqref{fig: H_variance}.

To proceed with the entanglement calculations, we shift our focus from exact energy diagonalization to the informational structure of the system. In continuous-variable systems, entanglement entropy is fundamentally governed by the covariance matrix (the second moments of the canonical operators). The hallmark of the IR divergence in standard quantum mechanics is the unbounded growth of the position variance, $\braket{\hat x^2} \to \infty$ as $\omega \to 0$. As established in Section~\eqref{section:SHO}, the primary geometric effect of the EUP is the strict saturation of this variance to $\braket{\hat x^2} \to 1/\gamma$.

Therefore, we propose a \emph{Moment-Matched Gaussian Reference State} to approximate the entanglement dynamics. While this Gaussian ansatz is not an exact energy eigenstate --- yielding relative energy fluctuations $\sigma_H / \braket{\hat H} \sim \mathcal{O}(1)$ due to the truncation of the non-Gaussian EUP tails --- it perfectly captures the regularized covariance structure of the system. By strictly enforcing the geometric bounds on spatial correlations dictated by the EUP, this moment-matched state allows us to cleanly isolate how the geometric cutoff of large-scale fluctuations functionally resolves the IR divergence of entanglement entropy, independent of higher-order non-Gaussian energy corrections. 

The use of a Gaussian Reference State (or Gaussian Equivalent State) is a mathematically robust and widely established technique in continuous-variable quantum information for approximating the entanglement dynamics of non-Gaussian systems~\cite{Genoni:2008zxa, Adesso:2014npz}. In this formal framework, the Gaussian reference state $\rho_G$ is uniquely defined as the specific Gaussian state that shares the exact same covariance matrix (first and second moments) as the true state $\rho$ in the physical regime of interest. According to the \emph{Maximum Entropy Principle}, for a given fixed covariance matrix, the corresponding Gaussian state strictly maximizes the von Neumann entropy~\cite{Genoni:2008zxa, DePalma:2021ekj}. Therefore, the entanglement entropy of our moment-matched Gaussian ansatz provides a rigorous upper bound to the true entanglement entropy of the system: $S(\rho_G) \geq S(\rho)$. By demonstrating that $S(\rho_G)$ saturates to a finite value in the IR limit ($\omega \to 0$) due to the EUP-imposed bounds on spatial variance, we formally prove that the true EUP entanglement entropy is fundamentally shielded from the standard IR divergence.

\subsection{Entanglement Entropy of the Gaussian Reference State}
\label{subsection:GaussianCHO}

A widely established approach for evaluating entanglement entropy in continuous-variable systems relies on computing the symplectic eigenvalues of the reduced density matrix (RDM) \cite{Adesso:2007jg,Adesso:2014npz,mallayadiv,chandran_one--one_2020, aimetHowCompactnessCurbs2026}. RDM is constructed by tracing out the degrees of freedom of the environmental subsystem from the global pure state of the system. RDM formalism evaluates entanglement exclusively from the structural configuration of the system's global wavefunction. Because the mathematical structure of a true wavefunction is fundamentally dictated by both the kinetic and potential operators within the governing Schr\"odinger equation, it encodes all requisite physical observables ---such as position and momentum variances --- necessary to determine the exact entanglement entropy.

Consequently, when evaluating entanglement dynamics via a variational or approximate state, it is imperative that the chosen trial wavefunction faithfully reproduces these exact information structures, at least in the asymptotic limits. The Gaussian Reference State (GRS) directly encounters a limitation in this regard~\cite{Genoni:2008zxa,Adesso:2014npz}. While GRS is deliberately constructed to exactly reproduce the EUP position variance, it cannot simultaneously accommodate the exact EUP momentum variance within the same pure-state wavefunctional structure. In fact, an explicit evaluation of the momentum operator on the GRS yields a momentum variance that deviates from the true EUP momentum variance. The specific physical regimes under which these momentum variances asymptotically converge are discussed in detail in Appendix~\eqref{appendix: mom_var_grf}.

Given that the RDM method relies strictly on the explicit functional form of the wavefunction, the GRS can be conceptually modeled as an \emph{exact eigenstate} of an effective substitute Hamiltonian. This effective system features a standard HUP kinetic operator, given by $-\frac{\hbar^2}{2M} \frac{d^2}{dx^2}$, coupled with a modified quadratic potential wherein the oscillator frequency is artificially deformed to match the EUP-defined spatial variance. An example of such an effective Hamiltonian is given by:
\begin{equation}
    \hat{H}_{\text{eff}} = \frac{\hat{p}^2}{2M} + \frac{1}{2}M\omega'^2 \hat{x}^2,
\end{equation}
which admits the exact ground-state solution $\psi(x) = \left(\frac{\beta}{\pi}\right)^{1/4} \exp\left(-\frac{\beta x^2}{2}\right)$, where $\hat{p} = -i\hbar \frac{d}{dx}$ and the effective frequency is fixed such that $\beta = \frac{M\omega'}{\hbar} = \beta_{\text{EUP}} = \frac{\gamma}{2}(2l+1)$.

Although this wavefunction precisely replicates the desired EUP position variance, the underlying physical mechanism is effectively inverted: the saturation of the position variance, which physically originates from the geometric modification of the kinetic sector in the true EUP framework, is algebraically mimicked here by a rigid, deformed potential sector. Because the RDM acts directly on this pure state, it fundamentally computes the entropy associated with this effective, potential-deformed HUP system rather than a system governed by a genuinely deformed EUP momentum operator. This distinction becomes acutely apparent when comparing the entanglement entropy profiles generated via the RDM method against those calculated using the Reduced Covariance Matrix (RCM) framework. Unlike the RDM, the RCM formalism is not rigidly constrained to a single pure-state wavefunction; rather, it allows for the manual, independent insertion of both the exact EUP momentum and position variances, directly capturing the full geometric deformation. This methodological comparison is illustrated in Appendix~\eqref{appendix: rcm_rdm_entanglement}.

Nevertheless, the GRS momentum variances evaluated across different frameworks --- namely $\braket{\hat p^2}_{\text{G, HUP}}$ and $\braket{\hat p^2}_{\text{G, EUP}}$ --- rigorously converge toward the exact EUP momentum variance $\braket{\hat p^2}_{\text{EUP}}$ in the deep asymptotic limit where $\kappa = \frac{M\omega_{\text{int}}}{\hbar \gamma} \gg 1$. For this reason, we deliberately omit the regime of ultra-small $\kappa$ values from our primary analysis. While energy fluctuations grow at large $\kappa$ scales, they remain systematically bounded within the same order of magnitude; concurrently, the residual coupling strength escalates. This unwanted growth can be successfully suppressed by keeping the internal interaction frequency $\omega_{\text{int}}$ fixed while driving the geometric deformation parameter $\gamma$ toward zero, as the scaling prefactor $\frac{\hbar^2\gamma}{M}$ effectively regularizes the overall increase. For robust numerical simulations, an optimal parameter range for $\kappa$ is found to lie strictly between $10^2$ and $10^4$.

Leveraging this regime, we directly utilize the standard Gaussian RDM solutions derived in Refs.~\cite{chandran_divergence_2019,aimetHowCompactnessCurbs2026} to compute the entropy:
\begin{equation}
    S(\xi)=-\ln (1-\xi) - \frac{\xi}{1-\xi}\ln \xi
\end{equation}
The parameter $\xi$ is defined entirely by the components of the covariance matrix:
\begin{align}
    \xi&= \frac{\gamma_2}{\gamma_1+ \varrho} \ ;\quad
    \varrho= \sqrt{\beta_+ \beta_-},\\
    \gamma_1&=\frac{\beta_+^2 +\beta_-^2 + 6\beta_+\beta_-}{4(\beta_+ + \beta_-)} ; \quad
    \gamma_2=\frac{(\beta_+ -\beta_-)^2}{4(\beta_+ + \beta_-)}
\end{align}

\begin{figure}[h]
    \centering
    \includegraphics[width=0.32\linewidth]{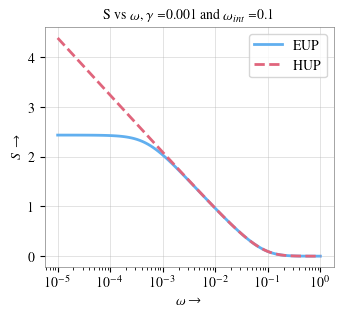}
    \includegraphics[width=0.32\linewidth]{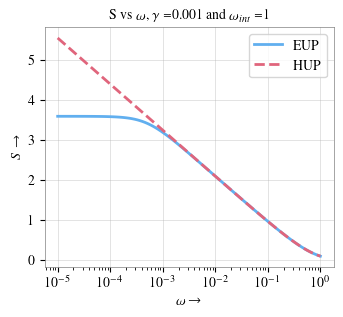}
    \includegraphics[width=0.32\linewidth]{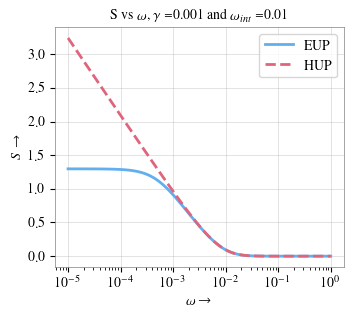}
    \caption{The entanglement entropy $S$ as a function of the external frequency $\omega$ for varying interaction strengths $\omega_{\text{int}}/\gamma$, calculated using natural units ($M=\hbar=1$).}
    \label{fig:SvswCoupled}
\end{figure}

In Fig.~\eqref{fig:SvswCoupled}, one can clearly observe that the low-frequency behavior is qualitatively different from its standard HUP counterpart. The geometric confinement inherent in the EUP framework strictly prevents the entanglement entropy from diverging. The transition to this saturated phase initiates at a critical frequency scale of $\omega \sim \hbar\gamma/M$.

A highly advantageous property of the moment-matched Gaussian ansatz is its direct extensibility to systems with macroscopically large degrees of freedom. In the following section, we extend this analytical framework to a one-dimensional chain of interacting harmonic oscillators characterized by an on-site potential and nearest-neighbor couplings.

\section{One-Dimensional Harmonic Chain}
\label{section:Chain}

Let us consider a one-dimensional chain consisting of $N$ identical masses ($M$ and frequency $\omega$) interacting via a nearest-neighbor quadratic potential, with each mass subject to the same on-site potential. The displacements of the sites are denoted by $\{x_i\}$ where $i=0,1,\dots,N-1$. For mathematical simplicity, we impose periodic boundary conditions (PBC), meaning $i=0$ and $i=N$ refer to the exact same mass~\cite{chandran_one--one_2020,Jain:2021ppx}. The Hamiltonian of the system is given by:
\begin{equation}
    \hat H = \sum_{i=0}^{N-1} \frac{\hat p_i^2}{2M} + \sum_{i=0}^{N-1} \frac{1}{2}M \omega^2\hat x_i^2 + \sum_{i=0}^{N-1} \frac{1}{2}M\omega_{\text{int}}^2(\hat x_{i+1}-\hat x_i)^2
    \label{eqn:chain}
\end{equation}
The standard procedure to solve such a Hamiltonian is to diagonalize the quadratic form in the potential sector. In the present setup, the Hamiltonian can be rewritten in matrix form as:
\begin{equation}
    \hat H = \sum_{i=0}^{N-1} \frac{\hat p_i^2}{2M} + \frac{M}{2}\sum_{i=0}^{N-1}\sum_{j=0}^{N-1} \hat x_i K_{ij}\hat x_j
\end{equation}
The eigenvalues and eigenvectors of the coupling matrix $\mathbf{K}$ correspond to discrete sine and cosine modes. For a chain with periodic boundary conditions, the normal-mode frequencies are explicitly given by:
\begin{equation}
    \tilde\omega_k^2 = \omega^2+4\omega_{\text{int}}^2\sin^2\left(\frac{\pi k}{N}\right) \;, \quad k \in \{0, 1, \dots, N-1\}
\end{equation}

To calculate the entanglement entropy, we utilize the reduced covariance matrix (RCM) approach discussed in Ref.~\cite{mallayadiv}. We define the reduced covariance matrix by considering the first oscillator as our primary subsystem and tracing out the remaining $N-1$ oscillators. This specific choice allows us to proceed with the majority of the calculations analytically. Because the Hamiltonian possesses time-reversal symmetry (in both the standard and EUP frameworks), the off-diagonal cross-correlator elements identically vanish. The reduced covariance matrix is thus defined as:
\begin{equation}
    \rho_{\text{red}} = \begin{bmatrix}
        \braket{\hat x_0^2} & 0\\
        0 & \braket{\hat p_0^2}
    \end{bmatrix}
\end{equation}
Note that the physical spatial variance is related to the normal-mode variances via:
\begin{equation}
    \braket{\hat x_i^2} = \sum_k U_{ik}^2\braket{\tilde{x}_k^2},
\end{equation}
where $\tilde{x}_k$ denotes the normal-mode coordinates. Under PBC, the transformation matrix elements satisfy $U_{ik}^2 = 1/N$. To approximately incorporate the EUP geometric effects into the many-body dynamics, we replace the standard normal-mode variances $\braket{\tilde{x}_k^2}$ and $\braket{\tilde{p}_k^2}$ with their exact EUP-modified counterparts.

However, the covariance matrix approach is strictly derived for Gaussian states, which are completely defined by their second moments. {In addition to the aforementioned consistency challenges between the RCM and RDM approaches, the RCM framework imposes a strict physical constraint requiring its symplectic eigenvalues to be strictly greater than or equal to $1/2$. This requirement is deeply rooted in the foundational structure of HUP, which dictates a minimum uncertainty relation of the form $\sigma_x \sigma_p \ge \hbar/2$. While the deep asymptotic regime ($\kappa = M\omega_{\text{int}}/\hbar\gamma \gg 1$) naturally guarantees the validity of this minimum uncertainty relation, a severe violation can occur in the opposite limit where $\kappa \ll 1$. To preserve physical consistency within this low-$\kappa$ domain, one must explicitly invoke the standard minimum uncertainty relation to reconstruct a valid momentum variance; that is, the HUP-based definition of momentum must be utilized. Consequently, the naive application of the RCM methodology becomes questionable within this specific extreme parameter regime, further justifying our focus on the physically well-behaved $\kappa \gg 1$ domain.} 

\begin{figure}[h]
    \centering
    \includegraphics[width=0.40\linewidth]{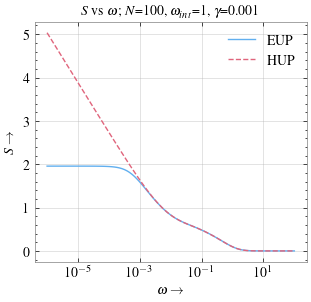}
    \includegraphics[width=0.40\linewidth]{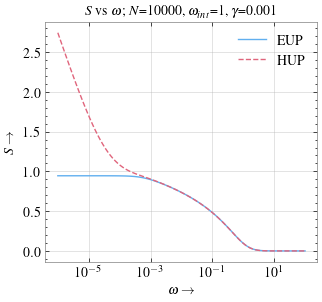}
    \caption{The entanglement entropy $S$ as a function of the on-site frequency $\omega$ for different chain lengths $N$, plotted using natural units ($M=\hbar=1$).}
    \label{fig:SvswChain}
\end{figure}

The resulting entanglement plots (Fig.~\eqref{fig:SvswChain}) are characteristically similar to those obtained for the two-oscillator case.
The qualitative departure from the standard HUP behavior consistently initiates near $\omega \sim \gamma$. This is not a numerical coincidence, but rather a direct consequence of the physical fact that the spatial variance saturates exactly at $1/\gamma$ in the zero-frequency (free-particle) limit. Thus, the dimensionless ratio $M\omega/\hbar\gamma$ acts as the fundamental parameter separating the geometric EUP regime from the standard HUP regime. {An approximate analytical value of entanglement entropy at zero frequency is derived for the continuum limit $N\to \infty$ at Appendix \eqref{appendix: contLimitChain}}.

\section{Scalar field theory in $(1+1)-$D within the EUP framework}
\label{section: fieldTheoryEUP}

Massless scalar fields exhibit IR divergences analogous to those of discrete harmonic chains~\cite{mallayadiv}. However, because the fundamental degrees of freedom in field theory are continuous variables defined over spacetime, the modified position--momentum algebra cannot be mapped directly onto the field operators. To formulate the scalar field theory under the EUP deformation, we need to derive the modified Klein-Gordon equation using the EUP momentum operator. Following the procedure established for the Schr\"odinger equation in Sec.~\eqref{sec:EUP}, we define the deformed spatial derivative operator as:
\begin{equation}
-i\hat\D \equiv -i\hbar(1+\gamma\hat{x}^2)\partial_x - i\hbar\gamma\hat{x} , . \label{eqn:mod_p_field}
\end{equation}
Working in natural units ($\hbar=c=1$), this operator simplifies to:
\begin{equation}
\hat{\D} = (1+\gamma\hat{x}^2)\partial_x + \gamma\hat{x} , .
\end{equation}
Like in the harmonic oscillator case, the transition from the HUP to the EUP framework is formally implemented via the spatial derivative mapping $\partial_x \to \hat{\D}$. Denoting the field mass as $m_f$, the action yields the modified Lagrangian:
\begin{align}
L &= \int \mathcal{L}\left(\Phi, \partial_x\Phi, \partial_t\Phi, t\right) dx \nonumber \\
&= \frac{1}{2}\int dx, \left[ (\partial_t\Phi)^2 - (1+\gamma x^2)^2(\partial_x\Phi)^2 - \left(m_f^2 - \gamma(1+2\gamma x^2)\right)\Phi^2 \right] .
\end{align}
Integrating the spatial derivative term by parts under vanishing boundary conditions leads to the following form:
\begin{equation}
L = \frac{1}{2}\int dx, \left[ (\partial_t\Phi)^2 - \Phi \hat{\D}^2 \Phi - m_f^2\Phi^2 \right] .
\end{equation}
Applying the Euler-Lagrange equations to the above modified Lagrangian density leads to the EUP-deformed Klein-Gordon equation:
\begin{equation}
(\hat{\D}^2 - \partial_t^2 - m_f^2)\Phi(x,t) = 0 \, . \label{eqn:KGeqn}
\end{equation}
Using the definition of the canonical momentum,
\begin{equation}
\Pi(x,t) = \frac{\partial\mathcal{L}}{\partial\dot{\Phi}} = \dot{\Phi}(x,t) \, ,
\end{equation}
the field Hamiltonian takes the following form:
\begin{equation}
H = \frac{1}{2}\int dx \left[ \Pi^2 + m_f^2\Phi^2 - \Phi \hat{\D}^2 \Phi \right] . \label{eqn: ScalarfieldHamiltonian}
\end{equation}
As demonstrated in Sec.~\eqref{section:SHO} for the simple harmonic oscillator case, the eigenvalue problem for $\hat{\D}^2$ is solved exactly by setting $l=0$ in Eq.~\eqref{eqn:eigstates}, which is equivalent to setting $\omega=0$ or $V(x)=0$ in Eq.~\eqref{eqn:SHO1}. The corresponding eigenfunctions satisfy:
\begin{equation}
\hat{\D}^2\varphi_n(x) = -\lambda_n^2 \varphi_n(x) , ,
\end{equation}
where $\lambda_n^2 = \gamma (n+1)^2$, and the set $\{\varphi_n(x)\}$ forms a complete orthonormal basis representing the EUP counterpart to the free-particle problem. As discussed in Sec.~\eqref{subsection: freeparticle-limit}, a discrete spectrum persists even in the absence of an explicit potential under the EUP. Consequently, rather than admitting the continuum Fourier basis of conventional free scalar fields, the EUP framework naturally dictates a discrete mode expansion:
\begin{align}
\hat{\Phi}(x,t) &= \sum_{n=0}^\infty \frac{1}{\sqrt{2\zeta_n}} \left[ \varphi_n(x) e^{-i\zeta_n t} \hat{a}_n + \varphi_n^*(x) e^{i\zeta_n t} \hat{a}_n^\dagger \right] , \\
\hat{\Pi}(x,t) &= -i \sum_{n=0}^{\infty} \sqrt{\frac{\zeta_n}{2}} \left[ \varphi_n(x) e^{-i\zeta_n t} \hat{a}_n - \varphi_n^*(x) e^{i\zeta_n t} \hat{a}_n^\dagger \right] ,
\end{align}
where $\zeta_n^2 = \lambda_n^2 + m_f^2$, and $\hat{a}_n, \hat{a}_n^\dagger$ are the annihilation and creation operators satisfying $[\hat{a}_m, \hat{a}_n^\dagger] = \delta_{mn}$ (see Appendix~\eqref{appendix: scalarfield} for detailed derivations). Furthermore, the field operators satisfy the standard equal-time commutation relation:
\begin{equation}
[\hat{\Phi}(x,t), \hat{\Pi}(y,t)] = i \delta(x-y) , .
\end{equation}
The mode-expanded normal-ordered Hamiltonian retains the standard harmonic structure:
\begin{equation}
:\hat{H}: = \sum_{n=0}^\infty \zeta_n \left( \hat{a}_n^\dagger \hat{a}_n + \frac{1}{2} \right) .
\end{equation}
This above result yields three immediate physical consequences:
\begin{enumerate}
\item \textbf{Discretization of the continuum spectrum:} In contrast to standard field theory in infinite flat space --- where momentum forms a continuous spectrum and fields are expanded in integrals over Fourier modes ($\int dk$) --- the EUP framework enforces a discrete mode sum ($\sum_n$). The continuous wave number $k$ is replaced by quantized spatial eigenvalues $\lambda_n = \sqrt{\gamma}(n+1)$, indexed by the integer mode number $n$.
\item \textbf{Curvature-induced mass gap:} The modified dispersion relation $\zeta_n = \sqrt{\gamma(n+1)^2 + m_f^2}$ implies that free quantum field excitations acquire a discrete, non-equidistant spectrum. Even in the strictly massless limit ($m_f = 0$), the ground-state field mode exhibits a non-zero energy gap $\zeta_0 = \sqrt{\gamma}$.
\item \textbf{Preservation of canonical quantization:} Since the equal-time commutation relation $[\hat{\Phi}(x,t), \hat{\Pi}(y,t)] = i\delta(x-y)$ remains local and canonical, the EUP deformation alters only the spatial mode profiles $\varphi_n(x)$ and energy levels $\zeta_n$, leaving the fundamental local quantum mechanical structure of the field intact.
\end{enumerate}

\subsection{Entanglement entropy of the EUP-modified scalar field}

To evaluate the entanglement entropy, we express the Hamiltonian in Eq.~\eqref{eqn: ScalarfieldHamiltonian} in the following self-adjoint form:
\begin{equation}
\hat{H} = \frac{1}{2}\int dx, \left[ \hat{\Pi}^2 + m_f^2\Phi^2 +\Phi \hat P^\dagger \hat P \Phi   \right] .
\end{equation}
Here $\hat P=-i\hat\D$. Discretizing the spatial coordinate $x$ onto a lattice with spacing $a$, the continuous equal-time commutator transforms from a Dirac delta function to a scaled Kronecker delta:
\begin{equation}
[\hat{\Phi}_i, \hat{\Pi}_j] = \frac{i}{a}\delta_{ij} , .
\end{equation}
Rescaling the field operators as $q_i = \sqrt{a}\,\hat{\Phi}_i$ and $p_i = \hat{\Pi}_i/\sqrt{a}$ restores canonical commutation relations, $[q_i, p_j] = i\delta_{ij}$. The discretized Hamiltonian then reads:
\begin{equation}
H = \frac{1}{2}\sum_{i=0}^N p_i^2 + \frac{1}{2}\sum_{i,j=0}^N q_i K_{ij} q_j \, .
\end{equation}
where $K_{ij}$ is the interaction matrix; as discussed below, the values of the interaction matrix depend on the choice of the boundary condition.
In the continuum limit $N \to \infty$ and $a \to 0$ (with total length $L = Na$ held fixed), this discrete system recovers the continuous field Hamiltonian in Eq.~\eqref{eqn: ScalarfieldHamiltonian}. We can consider two standard boundary conditions for this lattice system~\cite{Jain:2021ppx}:
\begin{itemize}
\item \textbf{Dirichlet boundary conditions:} Corresponding to a pinned chain of oscillators, the boundary fields are constrained to vanish ($q_{-1} = q_N = 0$). Because the endpoints are fixed, the center-of-mass mode is pinned, and the normal-mode frequencies remain strictly positive as $m_f \to 0$. Consequently, no zero mode exists under this boundary condition~\cite{Jain:2021ppx}.
\item \textbf{Neumann boundary conditions:} Equivalent to an open chain with free endpoints, Neumann boundary conditions enforce vanishing spatial derivatives at the boundaries ($\partial_x \Phi\big\vert{}_{\text{boundary}} = 0$). In the discrete domain, this translates to $q_0 = q_1$ and $q_N = q_{N-1}$. The unconstrained boundary oscillators allow a completely delocalized center-of-mass mode, yielding a zero mode as $m_f \to 0$ in the standard HUP framework~\cite{chandran_one--one_2020,Jain:2021ppx}.
\end{itemize}
Since Neumann boundary conditions host the zero mode governing IR behavior, we impose them to study the entanglement entropy. Under these conditions, we discretize the Hermitian operator $\hat P$
using a forward finite-difference scheme and construct the coupling matrix as:
\begin{equation}
K=P^\dagger P+m_f^2I.
\end{equation}
The resulting coupling matrix takes the explicit tridiagonal form:
\begin{equation}
K_{ij} = \delta_{ij} \left[ m_f^2 + \left( \frac{g_i}{a} - \gamma x_i \right)^2 + \frac{g_{i-1}^2}{a^2} \right] - \delta_{i,j-1} \frac{g_i}{a} \left( \frac{g_i}{a} - \gamma x_i \right) - \delta_{i-1,j} \frac{g_j}{a} \left( \frac{g_j}{a} - \gamma x_j \right),
\end{equation}
where
\begin{equation}
g_i = 1 + \gamma x_i^2.
\end{equation}
The behavior of the coupling matrix $K_{ij}$ across distinct physical regimes highlights the impact of the EUP deformation onto the quantum field:
\begin{itemize}

\item \textbf{Standard harmonic chain limit ($\gamma \to 0$):}
As $\gamma\to0$, $g_i\to1$, reducing $K_{ij}$ to the translationally invariant tridiagonal matrix:
\begin{equation}
K_{ij} = \delta_{ij} \left( m_f^2 + \frac{2}{a^2} \right) - \frac{\delta_{i,j-1} + \delta_{i-1,j}}{a^2},
\end{equation}
characterized by standard cosine modes. Under Neumann boundary conditions at $m_f=0$, the lowest eigenvalue approaches zero, recovering the uniform zero mode responsible for the standard IR divergence in the entanglement entropy.

\item \textbf{Massless limit and boundary asymptotics ($m_f\to0$, large $\vert x_i\vert$):}
Setting $m_f=0$ leaves a strictly positive, non-constant diagonal potential:
\begin{equation}
K_{ii} = \left( \frac{g_i}{a} - \gamma x_i \right)^2 + \frac{g_{i-1}^2}{a^2}.
\end{equation}
Unlike the flat-space limit, the lowest eigenvalue no longer vanishes due to the EUP deformation, manifesting as a non-vanishing spatial mass gap of order $\zeta_0\sim\sqrt{\gamma}$. Far from the origin ($\vert x_i\vert\gg\gamma^{-1/2}$), the asymptotic scaling $g_i\approx\gamma x_i^2$ causes the diagonal elements to grow quartically as:
\begin{equation}
K_{ii} \sim \frac{2\gamma^2x_i^4}{a^2},
\end{equation}
which acts as an effective confining potential near the boundaries, keeps $K_{ij}$ strictly positive-definite, and lifts the zero mode.

\item \textbf{Heavy field limit ($m_f^2\gg\gamma\,a^{-2}$):}
When the physical field mass dominates over both the lattice derivative and deformation scales:
\begin{equation}
K_{ij}\approx m_f^2\delta_{ij}.
\end{equation}
Nearest-neighbor couplings become negligible ($K_{i,i\pm1}\ll K_{ii}$), causing spatial correlations to decay rapidly over the correlation length $\xi\sim m_f^{-1}\to0$ and driving the entanglement entropy across any partition towards zero.

\end{itemize}

\begin{figure}[H]
\centering
\includegraphics[width=0.45\linewidth]{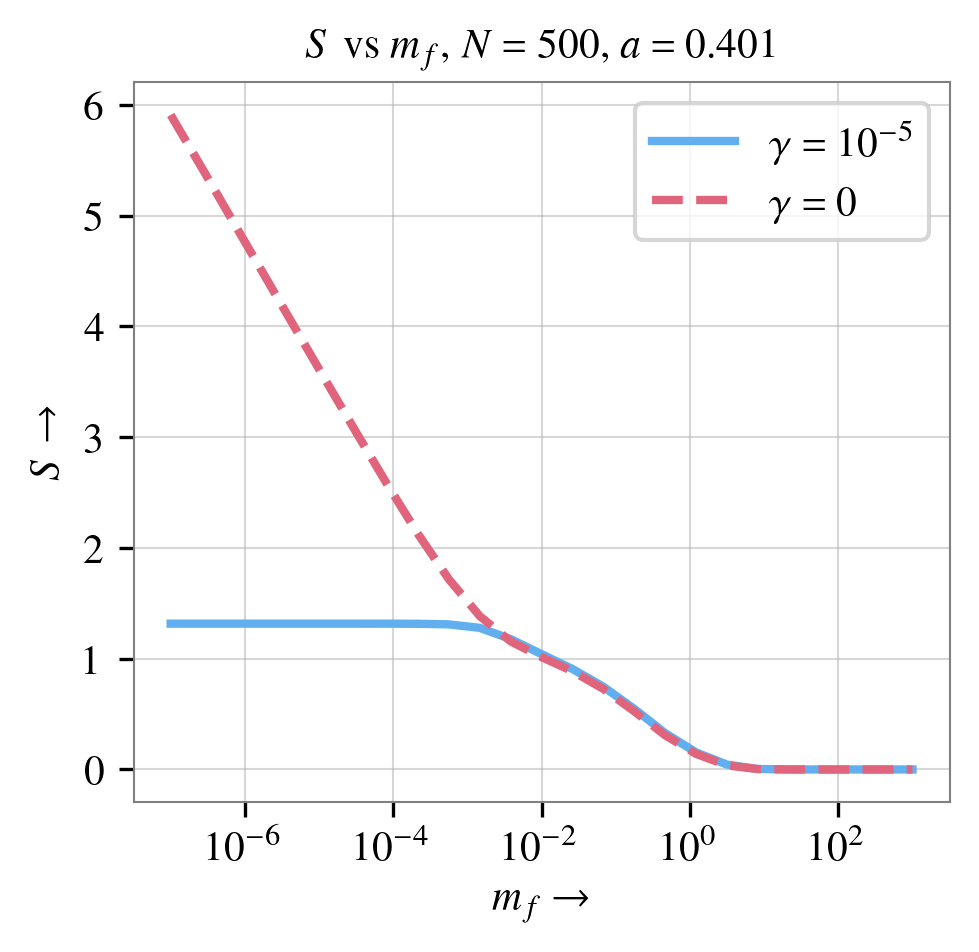}
\includegraphics[width=0.45\linewidth]{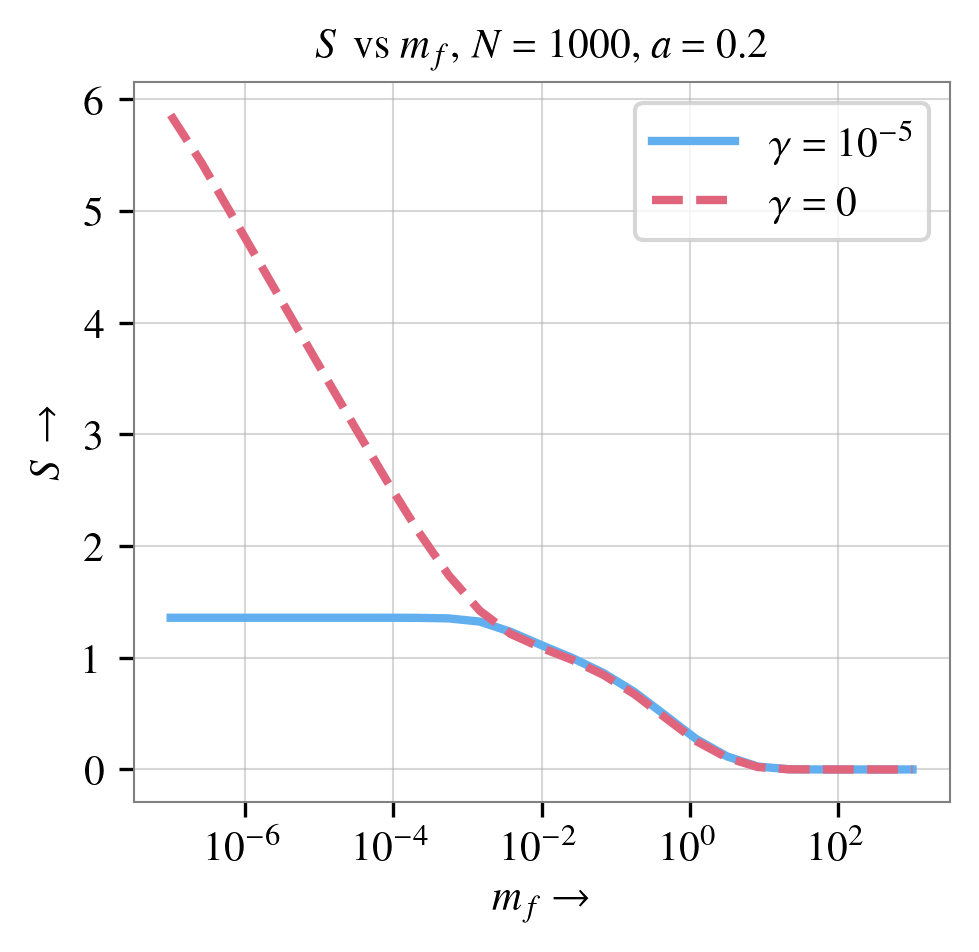}
\caption{Entanglement entropy $S$ as a function of the scalar field mass $m_f$.}
\label{fig:placeholder}
\end{figure}

We evaluate the entanglement entropy via the covariance matrix formalism~\cite{Bombelli:1986rw,Adesso:2007jg,Adesso:2014npz,mallayadiv,chandran_one--one_2020,Jain:2021ppx}. To go about that, we introduce a unitary transformation such that $K = U^\dagger D U$, where $D = \text{diag}\{\chi_i^2\}$ is diagonal. Defining the coordinate transformation $\tilde{q}_i = U_{ij}q_j$, the conjugate momenta transform as $\tilde{p}_i = (U^\dagger)_{ij} p_j$ because $p_i$ and $q_i$ obey standard canonical commutation relations. In this normal-mode basis, the Hamiltonian decouples:
\begin{equation}
H = \frac{1}{2} \sum_{i=0}^N \left(\tilde{p}_i^2 + \chi_i^2\tilde{q}_i^2\right) .
\end{equation}
Defining the frequency matrix $\Omega = D^{1/2}$, the global covariance matrix is given by:
\begin{equation}
\sigma = \begin{pmatrix}
X & 0 \
0 & P
\end{pmatrix} ,
\end{equation}
where $X_{ij} = \braket{q_iq_j} = (U^\dagger\Omega^{-1} U)_{ij}/2$ and $P_{ij} = \braket{p_ip_j} = (U^\dagger\Omega U)_{ij}/2$. To construct the reduced covariance matrix, we trace out the environment and retain only the first $N_{\text{sub}}$ oscillators, denoting the reduced blocks as $X_{\text{sub}}$ and $P_{\text{sub}}$. The symplectic eigenvalues $\{\nu_i\}$ correspond to the standard eigenvalues of the matrix $\sqrt{X_{\text{sub}} P_{\text{sub}}}$. Finally, the von Neumann entanglement entropy is evaluated as:
\begin{equation}
S = \sum_{i=0}^{N_{\text{sub}}} \left[ \left(\nu_i+\frac{1}{2}\right)\ln!\left(\nu_i+\frac{1}{2}\right) - \left(\nu_i-\frac{1}{2}\right)\ln!\left(\nu_i-\frac{1}{2}\right) \right] .
\end{equation}
The numerical results presented in Fig.~\eqref{fig:placeholder} exhibit behavior analogous to the one-dimensional oscillator chain: rather than displaying the standard IR divergence as $m_f \to 0$, the entanglement entropy saturates to a finite plateau. This behavior is reminiscent of the finite zero-point energy of free particles in the EUP framework. This saturation is directly attributable to the emergence of an effective geometric mass contribution within the coupling matrix $K_{ij}$. Even in the strictly massless limit ($m_f = 0$), the diagonal elements $K_{ii}$ grow quadratically with position $x_i$ due to the spatial dependence of $f_i$. Consequently, for a symmetric spatial domain, the diagonal entries reach a minimum at the origin and grow toward the boundaries, establishing an effective spatial confinement.

A crossover in the scaling of the entanglement entropy occurs when the explicit field mass $m_f$ competes with and dominates this geometric mass contribution. Dimensional analysis implies that this transition point scales as $m_f \sim \sqrt{\gamma}$ in natural units or $m_f \sim {\hbar}\sqrt{\gamma}/c$ in SI units. This threshold scale coincides precisely with the zero-point energy scale of a free particle in the EUP framework.

\subsection{Entanglement Spectrum}

While the von Neumann entanglement entropy provides a single scalar measure of the overall bipartite quantum correlation across a spatial partition, it fundamentally represents a thermodynamic average that compresses the informational structure of the vacuum state. A more complete and fine-grained characterization is encoded within the entanglement spectrum~\cite{Li:2008kda,Chandran:2013zqa,2017-Kumar.Shankaranarayanan-SRep,Jain:2021ppx}. Formally defined via the spectrum of the modular (or entanglement) Hamiltonian $\hat{H}_E = -\ln \hat{\rho}_{\text{red}}$, the entanglement spectrum resolves the full hierarchical distribution of the Schmidt weights. In continuous-variable quantum field theories, the low-lying eigenvalues of this operator govern the long-range correlation physics and dictate whether the spatial subsystem exhibits effective thermalization, critical scaling, or topological order. Consequently, analyzing how these modular eigenvalues respond to geometric deformations allows us to isolate precisely how the EUP modifies the underlying informational architecture of the vacuum.

Since the reduced covariance matrix (RCM) method captures the exact Gaussian entanglement dynamics prescribed by the reduced density matrix (RDM), the complete entanglement spectrum can be extracted analytically directly from the symplectic eigenvalues. To keep the analytical derivation tractable while retaining the essential spectral features of the field, we restrict our primary subsystem to a single spatial lattice site ($N_{\text{sub}} = 1$). By equating the von Neumann entropy evaluated via the RDM eigenvalues $\{p_n\}$ with the symplectic representation of the RCM, we obtain:
\begin{equation}
S = -\sum_{n=0}^\infty p_n \ln p_n =
\left(\nu+\frac{1}{2}\right)\ln\left(\nu+\frac{1}{2}\right)
- \left(\nu-\frac{1}{2}\right)\ln\left(\nu-\frac{1}{2}\right) , \label{eqn: entropyRCMRDM}
\end{equation}
where $\nu$ represents the symplectic eigenvalue of the reduced covariance matrix. The functional identity in Eq.~\eqref{eqn: entropyRCMRDM} strictly demands that the RDM eigenvalues follow the thermal-like geometric distribution:
\begin{equation}
p_n = \frac{1}{\nu+1/2}\left(\frac{\nu - 1/2}{\nu + 1/2}\right)^n .
\end{equation}
Taking the negative logarithm of the reduced density matrix eigenvalues yields the exact discrete eigenvalues of the entanglement Hamiltonian, which constitute the entanglement spectrum:
\begin{equation}
h_n = -\ln p_n = \ln\left(\nu+\frac{1}{2}\right) + n\ln{\left(\frac{\nu + 1/2}{\nu - 1/2}\right)} . \label{eqn: ent_spectrum_levels}
\end{equation}

In the strictly massless limit ($m_f = 0$) subjected to a finite geometric deformation ($\gamma > 0$), the symplectic eigenvalue $\nu$ saturates to a constant, finite upper bound. Consequently, the entire entanglement spectrum in Eq.~\eqref{eqn: ent_spectrum_levels} reduces to a linear function of the mode index $n$. This behavior departs from the standard HUP framework, wherein the spectrum collapses toward a single, infinitely dense point as $m_f \to 0$ and $\nu \to \infty$. In the EUP framework, the spectrum remains discrete and evenly spaced even at absolute zero field mass. This linear dependence indicates that the entanglement spectrum shares the structural distribution of a thermal harmonic oscillator, with the saturated symplectic eigenvalue $\nu$ establishing an effective excitation gap. The parameter $\nu$ thus governs the eigenvalue spacing and can be interpreted as setting a finite effective entanglement temperature for the subsystem.

\begin{figure}[!h]
\centering
\includegraphics[width=0.45\linewidth]{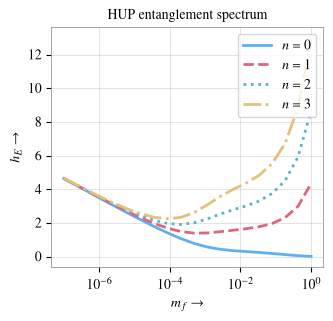}
\includegraphics[width=0.45\linewidth]{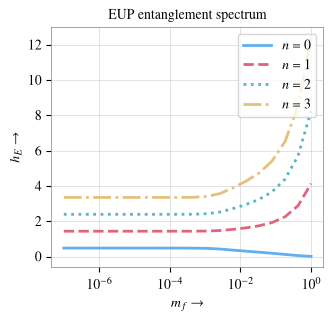}
\caption{The entanglement spectrum for the standard HUP (left) and EUP-modified (right) scalar fields, evaluated for parameters $N=200$, $\gamma=10^{-5}$, and $a=1$.}
\label{fig:EEspectrum}
\end{figure}

The closing of the entanglement gap in the standard HUP framework produces a continuous, un-gapped spectrum. This accumulation of low-lying modular modes is directly responsible for the IR divergence of the entanglement entropy in massless theories. In contrast, the EUP framework structurally enforces a constant, non-vanishing entanglement gap at low frequencies. This target-space geometric deformation acts as an energetic barrier in the modular flow that suppresses the proliferation of low-energy entanglement modes, thereby guaranteeing a finite entanglement entropy. The comparative structure of this regularized spectrum is illustrated in Fig.~\eqref{fig:EEspectrum}.

The transition of the entanglement spectrum from a collapsing continuous limit in the HUP framework to a discrete, evenly spaced configuration in the EUP framework indicates structural modifications in the entanglement Hamiltonian $\hat{H}_E$. Algebraically, the target-space curvature parameter $\gamma$ prevents the closing of the modular spectral gap ($\epsilon_{\text{min}} > 0$), which regularizes the effective entanglement temperature of the subsystem ($T_{\text{eff}} \le T_{\text{eff}}^{\text{max}}$). From an information-theoretic perspective, this demonstrates that field-amplitude saturation restricts the infinite correlation storage capacity typically associated with un-gapped zero modes. This behavior provides a field-theoretic realization of a dynamic IR screening mechanism, the broader implications of which are currently under investigation.

\section{Conclusions}

In this work, we have demonstrated that EUP provides a natural, fundamental resolution to the IR divergences of entanglement entropy in continuous-variable quantum systems. By introducing a geometric modification to the canonical commutation relations at macroscopic scales, the EUP fundamentally alters the low-energy dynamics of quantum fields. 

Our analysis of the simple harmonic oscillator in Section~\eqref{section:SHO} establishes that the EUP introduces an intrinsic spatial confinement, governed by a characteristic length scale of $\sim\gamma^{-1/2}$. This coordinate deformation acts as a geometric effective potential, restricting the wavefunction from spreading over infinite spatial extents and yielding a discrete, particle-in-a-box-like spectrum even in the free-particle limit ($\omega \to 0$). By placing a strict upper bound on the position variance, the EUP effectively excises the unbounded, arbitrarily long-wavelength fluctuations that characterize zero-modes. In the flat-space limit ($\gamma \to 0$), this geometric confinement vanishes, restoring continuous free-particle spectra and full translational invariance. 

A central methodological result of this work is the rigorous extension of these single-particle geometric bounds to interacting many-body systems, as detailed for coupled oscillators (Section~\eqref{section:CHO}) and 1D harmonic chains (Section~\eqref{section:Chain}). To navigate the highly non-Gaussian power-law tails of the exact EUP wavefunctions, we introduced a Moment-Matched Gaussian Reference State. By strictly enforcing the EUP-modified spatial variances within the reduced covariance matrix, we leveraged the Maximum Entropy Principle to establish a rigorous upper bound on the entanglement entropy. We explicitly demonstrated that the product state adheres to this maximum entropy bound in the specified $l_- \gg 1$ limit. Because this maximized Gaussian entropy saturates to a finite value, we mathematically prove the finiteness of the true entanglement entropy in the extreme IR limit.

Physically, the marked trend change in the entanglement entropy curve at $M\omega/\hbar \sim \gamma$ signals the emergence of a maximum quantum correlation length ($\sim\gamma^{-1/2}$). Beyond this geometric horizon, quantum correlations are exponentially suppressed. {We extended our analysis to $(1 + 1)-$D scalar field theory, considering EUP deformation, conjugate fields obey the standard equal time commutation relation. We explicitly calculated the mode expansion (as shown in Appendix~\eqref{appendix: scalarfield}) implying that the Hamiltonian also mimics the standard mode expansion. While the deformation arises in the spatial derivative of fields along with an extra geometric mass term, a unitary transformation is possible to diagonalize the Hamiltonian, effectively converting into decoupled harmonic oscillators. Nevertheless, the entanglement entropy saturates at the massless limit --- analogous to coupled oscillators discussed in Sec.~\eqref{section:CHO}}. Finite-system-size soft cutoff traditionally required to regularize field-theoretic IR divergences (e.g., in Ref.~\cite{mallayadiv}) emerges completely naturally from the EUP algebra. Consequently, the entanglement entropy exhibits a finite $\log(\log(16/\gamma))$ dependence, replacing the pathological divergence with a parameter entirely determined by the background geometry.

Ultimately, the underlying physical mechanism driving this saturation is revealed within the entanglement spectrum of the reduced density matrix. In standard quantum mechanics, the entanglement gap closes in the low-frequency limit due to the unchecked proliferation of zero-modes, resulting in an infinitely dense spectrum and divergent entropy. In stark contrast, the EUP protects this gap. By enforcing a finite spatial spread, the EUP suppresses the thermodynamic accumulation of IR modes, ensuring a non-vanishing entanglement gap that fundamentally shields the field from the IR catastrophe. These findings suggest that macroscopic spacetime geometry intrinsically regularizes quantum information, offering a robust framework for studying entanglement in cosmological and curved spacetime scenarios without relying on artificial boundary conditions.

\vspace{0.5cm}
\noindent\emph{\underline{Acknowledgement:}} The authors are grateful to S.~Mahesh Chandran, P. G. Christopher, H.~Dhar, K.~Hari, and T.~Parvez for their valuable discussions and feedback on the earlier draft. The MHRD fellowship at IIT Bombay financially supports SM. The work is supported by ANRF-Advanced Research Grant (ANRF/ARG/2025/001514/PS).  

\appendix

\section{EUP-Deformed Simple Harmonic Oscillator}
\label{appendix:normalization}

In this appendix, we provide explicit mathematical derivations supporting the continuous and discrete quantum properties of SHO deformed by EUP. Specifically, we establish a closed-form expression for the normalization constant of an arbitrary oscillator eigenstate, calculate the exact ground-state position and momentum variances, and derive the free-particle limit ($\omega \to 0$) under both Dirichlet and Neumann boundary conditions in a box of finite size $2R$ before evaluating the thermodynamic limit $R \to \infty$.

\subsection{Normalization constant for arbitrary state}

The calculation of the normalization constant for an arbitrary state involves evaluating analytical integrals of hypergeometric functions. However, for the special case where $l$ and $\gamma$ are real, positive quantities, a distinct pattern emerges upon examining the first few eigenstates. Based on this structure, we conjecture the following closed-form expression for the norm of an arbitrary state:
\begin{equation}
    \braket{\psi_n|\psi_n} = \sqrt{\frac{\pi}{\gamma}} \left( \frac{n!}{2^n} \right) \left[ \frac{ \Gamma \left( l + \frac{3}{2} + \frac{n - \sin^2\left(\frac{n\pi}{2}\right)}{2} \right) \, \Gamma \left( l + 1 + \frac{n + \sin^2\left(\frac{n\pi}{2}\right)}{2} \right) }{ \Gamma(l+n+1) \, \Gamma(l+n+2) } \right]
\end{equation}

This expression has been verified symbolically up to $n=30$ using Wolfram Mathematica, and numerically up to $n=50$. Due to the highly oscillatory nature of the integrand, an exceptionally fine spatial grid is required to evaluate the integral accurately. Prior to numerical integration, one must ensure that the wave function is smoothly resolved across the computational domain. Furthermore, to prevent numerical overflow or underflow for large values of $l$, we computed the normalization factors via the logarithm of the Gamma function (e.g., \texttt{gammaln} in Python) before exponentiation.

\subsection{Ground-State Variances}
\label{appendix: vargsw}

The normalized ground-state wave function of the EUP-deformed harmonic oscillator is given by:
\begin{equation}
    \psi(x) = \left(\frac{\gamma}{\pi}\right)^{1/4} \left(\frac{\Gamma(l+2)}{\Gamma(l+3/2)}\right)^{1/2} \frac{1}{(1+\gamma x^2)^{1+l/2}} \, .
\end{equation}
Since, the wave function is symmetric under parity ($x \to -x$), the first moment vanishes identically, $\braket{\hat{x}} = 0$. Consequently, the spatial variance is determined entirely by the second moment $\braket{\hat{x}^2}$, evaluated via the definite integral:
\begin{equation}
    \int_{-\infty}^{\infty} \frac{x^{2}}{\left(1+\gamma x^{2}\right)^{2+l}} \, dx = \frac{\sqrt{\pi}}{2\,\gamma^{3/2}} \frac{\Gamma\!\left(l+\frac{1}{2}\right)}{\Gamma(l+2)} \, , \qquad l>0, \ \gamma>0 \, .
\end{equation}

Substituting this integral into the variance expression yields:
\begin{equation}
    \braket{\hat{x}^2} - \braket{\hat{x}}^2 = \int_{-\infty}^{\infty} x^2 \psi^*(x)\psi(x) \, dx = \frac{1}{\gamma(2l+1)} \, .
\end{equation}
In the IR limit ($\omega \to 0$, corresponding to $l \to 0$), the spatial variance saturates to $1/\gamma$.
Using the ground-state expectation value of the EUP Hamiltonian together with the algebraic relation connecting $M\omega$ and $l$, the momentum variance is found to be:
\begin{equation}
    \braket{\hat{p}^2} = \hbar^2\gamma \frac{(l+1)^2}{2l +1} \, .
\end{equation}

\subsection{Free-Particle Limit $(\omega \to 0)$ Under the EUP}
\label{appendix: free-particle-details}

Evaluating the free-particle limit ($\omega \to 0$) of the EUP-deformed oscillator provides the continuous spectrum regime under modified spatial kinematics. Unlike standard quantum mechanics --- where free particles correspond to unconstrained plane waves --- the spatial deformation parameter $\gamma$ alters both the differential operator and the asymptotic behavior at spatial infinity. To establish box-normalized wave functions and quantify how boundary conditions shape the momentum spectrum, we analyze the equation on a finite interval $x \in [-R, R]$ under Dirichlet and Neumann boundary conditions before taking $R \to \infty$.

\subsubsection{Hypergeometric Reduction and Normalization}

In the zero-frequency limit $\omega = 0$ (which sets $l = 0$ in the general oscillator wave equation, Eq.~\eqref{eqn: generalsoln}), the spatial equation simplifies via the Gauss hypergeometric identity ${}_2F_1(1, 0; c; z) = 1$ to
\begin{equation}
    \varphi(x) = \frac{N}{\sqrt{1+\gamma x^2}} \left( \frac{1 \pm i \sqrt{\gamma} x}{1 \mp i \sqrt{\gamma} x} \right)^{m/2} = \frac{N}{\sqrt{1 + \gamma x^2}} \exp\left[ \pm i m \arctan\left(\sqrt{\gamma} x\right) \right] .
\end{equation}
To determine the normalization constant $N$, we consider the symmetric domain $x \in [-R, R]$:
\begin{equation}
    N^{-2} = \int_{-R}^R \frac{dx}{1 + \gamma x^2} = \frac{2}{\sqrt{\gamma}} \arctan\left(\sqrt{\gamma} R\right) .
\end{equation}
In the infinite-box limit $R \to \infty$, $\arctan(\sqrt{\gamma}R) \to \pi/2$, yielding $N = \gamma^{1/4}/\sqrt{\pi}$.

\subsubsection{Dirichlet Boundary Conditions}

Enforcing Dirichlet boundary conditions $\psi(\pm R) = 0$ on the general linear combination of states (Eq.~\eqref{eqn: freeparticleGnrlEigenstatesMain}) yields a system of homogeneous linear equations for the expansion coefficients $A$ and $B$. Requiring a non-trivial solution sets the determinant to zero, giving the quantization condition:
\begin{equation}
    \exp\left[4i m \arctan\left(\sqrt{\gamma} R\right)\right] = 1 = e^{2in\pi} \qquad (n \in \mathbb{Z}) \, ,
\end{equation}
which fixes the parameter $m$ as:
\begin{equation}
    m = \frac{n\pi}{2\arctan\left(\sqrt{\gamma} R\right)} \, .
\end{equation}

The boundary conditions mandate $B = (-1)^{n+1}A$. The explicit finite-box wave functions then reduce to:
\begin{equation}
    \psi_n(x) = \begin{cases}
        \dfrac{\gamma^{1/4}}{\sqrt{\arctan\left(\sqrt{\gamma} R\right)(1+\gamma x^2)}} \cos\left[ \dfrac{n\pi \arctan\left(\sqrt{\gamma} x\right)}{2\arctan\left(\sqrt{\gamma} R\right)} \right] , & \text{if } n \text{ is odd,} \\[1.2em]
        \dfrac{i\gamma^{1/4}}{\sqrt{\arctan\left(\sqrt{\gamma} R\right)(1+\gamma x^2)}} \sin\left[ \dfrac{n\pi \arctan\left(\sqrt{\gamma} x\right)}{2\arctan\left(\sqrt{\gamma} R\right)} \right] , & \text{if } n \text{ is even.}
    \end{cases}
\end{equation}
Taking the limit $R \to \infty$ recovers the continuum expressions discussed in Sec.~\eqref{subsec: free-particle-eup}.

\subsubsection{Neumann Boundary Conditions}

Applying Neumann boundary conditions $\psi'(\pm R) = 0$ to Eq.~\eqref{eqn: freeparticleGnrlEigenstatesMain} leads to a transcendental equation for the allowed modes $m \equiv m_n$:
\begin{equation}
    m_n \arctan\left(\sqrt{\gamma} R\right) = \arctan\left( \frac{m_n}{\sqrt{\gamma} R} \right) + \frac{n\pi}{2} \qquad (n \in \mathbb{Z}) \, .
\end{equation}
In the thermodynamic limit $R \to \infty$, the term $\arctan(m_n / \sqrt{\gamma}R) \to 0$ while $\arctan(\sqrt{\gamma}R) \to \pi/2$, reducing this relation to integer quantization $m_n \to n$.

The resulting wave functions take the same functional form as the Dirichlet case with $n$ replaced by $m_n$:
\begin{equation}
    \psi_n(x) = \begin{cases}
        \dfrac{\gamma^{1/4}}{\sqrt{\arctan\left(\sqrt{\gamma} R\right)(1+\gamma x^2)}} \cos\left[ \dfrac{m_n\pi \arctan\left(\sqrt{\gamma} x\right)}{2\arctan\left(\sqrt{\gamma} R\right)} \right] , & \text{if } n \text{ is odd,} \\[1.2em]
        \dfrac{i\gamma^{1/4}}{\sqrt{\arctan\left(\sqrt{\gamma} R\right)(1+\gamma x^2)}} \sin\left[ \dfrac{m_n\pi \arctan\left(\sqrt{\gamma} x\right)}{2\arctan\left(\sqrt{\gamma} R\right)} \right] , & \text{if } n \text{ is even.}
    \end{cases}
\end{equation}
As $R \to \infty$, these wave functions smoothly converge to the continuum results outlined in Sec.~\eqref{subsec: free-particle-eup}.

\section{EUP-Deformed Coupled Harmonic Oscillators}
\label{appendix:CHO}

In this appendix, we provide detailed derivations demonstrating why the exact eigenstates of the decoupled Hamiltonian $H'$ fail as a reliable approximation for the physical Hamiltonian $H$. Furthermore, we derive the momentum variances for the Gaussian reference state under the EUP framework.

\subsection{Energy Variance of the Product State Ansatz}
\label{appendix: variance_product_state}

We explicitly demonstrate why the exact eigenstates of the decoupled Hamiltonian $H'$ fail as a reliable approximation for the physical Hamiltonian $H$. Let $\ket{\Psi_{\text{prod}}}$ denote the trial product state, defined as an exact eigenstate of the unperturbed Hamiltonian $H'$ with eigenvalue $E'$:
\begin{equation}
    H' \ket{\Psi_{\text{prod}}} = E' \ket{\Psi_{\text{prod}}} \, .
\end{equation}
The physical Hamiltonian is decomposed as $H = H' - \Delta H$. The expectation value of the total energy in this trial state is:
\begin{equation}
    \braket{H} = \braket{\Psi_{\text{prod}}| H' - \Delta H |\Psi_{\text{prod}}} = E' - \braket{\Delta H} \, .
\end{equation}
To assess the quantitative validity of this approximation, we evaluate the expectation value of the Hamiltonian squared:
\begin{equation}
    \braket{H^2} = \braket{(H' - \Delta H)^2} = \braket{(H')^2 - H'\Delta H - \Delta H H' + (\Delta H)^2} \, .
\end{equation}
Because $\ket{\Psi_{\text{prod}}}$ is an eigenstate of $H'$, the actions of $H'$ on the state vectors simplify to:
\begin{align}
    \braket{(H')^2} &= (E')^2 \, , \\
    \braket{H'\Delta H} &= E'\braket{\Delta H} \, , \\
    \braket{\Delta H H'} &= E'\braket{\Delta H} \, .
\end{align}
Substituting these relations into the expression for $\braket{H^2}$ yields:
\begin{equation}
    \braket{H^2} = (E')^2 - 2E'\braket{\Delta H} + \braket{(\Delta H)^2} \, .
\end{equation}
The energy variance $\sigma_H^2 = \braket{H^2} - \braket{H}^2$ of the trial state is therefore:
\begin{equation}
\begin{split}
    \sigma_H^2 &= \left[ (E')^2 - 2E'\braket{\Delta H} + \braket{(\Delta H)^2} \right] - \left[ E' - \braket{\Delta H} \right]^2 \\
    &= (E')^2 - 2E'\braket{\Delta H} + \braket{(\Delta H)^2} - \left( (E')^2 - 2E'\braket{\Delta H} + \braket{\Delta H}^2 \right) \\
    &= \braket{(\Delta H)^2} - \braket{\Delta H}^2 = \sigma_{\Delta H}^2 \, .
\end{split}
\end{equation}
Thus, the energy variance of the physical Hamiltonian reduces exactly to the variance of the interaction difference operator $\Delta H$. 

The physical consequence of this exact equivalence becomes evident upon examining the mathematical structure of $\Delta H$ and the asymptotic behavior of the EUP wave functions. The difference operator $\Delta H$ introduces non-linear coordinate-momentum couplings consisting of paired position operators and spatial derivatives (e.g., $x_i x_j \frac{\partial^2}{\partial x_k \partial x_l}$). Consequently, evaluating $\braket{(\Delta H)^2}$ requires computing higher-order differential combinations of the form $x^{2n}\frac{\partial^{2n}}{\partial x^{2n}}$.

As derived in Section~\eqref{section:SHO}, the exact wave functions in the EUP framework exhibit power-law asymptotic tails:
\begin{equation}
    \psi(x) \sim \frac{1}{(1+\gamma x^2)^{1+l/2}} \, .
\end{equation}
In the IR limit ($\omega \to 0$, corresponding to $l \to 0$), the action of a $2n$-th order spatial derivative on $\psi(x)$ increases the power of the denominator to $(1+\gamma x^2)^{1+n}$ while generating a polynomial of degree $2n$ in the numerator. When computing the expectation value $\braket{(\Delta H)^2}$, the resulting integrand scales asymptotically at spatial infinity according to:
\begin{equation}
    \mathcal{I}(x) \sim \psi(x) \left[ x^{2n} \frac{\partial^{2n}}{\partial x^{2n}} \psi(x) \right] \sim \frac{1}{1+\gamma x^2} \cdot \frac{x^{2n}}{(1+\gamma x^2)^{1+n}} \sim \frac{x^{2n}}{x^{4+2n}} = \frac{1}{x^4} \, .
\end{equation}
Because the integrand decays as $x^{-4}$, the individual expectation values remain strictly convergent and do not exhibit formal ultraviolet divergences. However, because a power-law tail decays significantly slower than an exponential tail, it lacks the strong asymptotic suppression required to bound the rapid combinatorial growth of the coefficients appearing in these higher-order differential terms. This absence of exponential damping allows $\braket{(\Delta H)^2}$---and consequently $\sigma_H^2$---to grow orders of magnitude larger than the expectation value of the energy itself ($\sigma_H \gg \braket{H}$). This mathematical behavior explains the large energy variance observed numerically and justifies the necessity of the Gaussian ansatz, which reintroduces an exponential tail to bind these higher-order fluctuations effectively.

\subsection{Momentum Variances of the Gaussian Reference State}
\label{appendix: mom_var_grf}

Consider a single-variable Gaussian state with an EUP-enforced position variance:
\begin{equation}
    \psi(x) = \left(\frac{\beta}{\pi}\right)^{1/4} \exp\left(-\frac{\beta x^2}{2}\right) ,
\end{equation}
where $\beta = \frac{\gamma}{2}(2l+1)$. In the standard HUP framework, the corresponding momentum variance is given by:
\begin{equation}
    \braket{\hat{p}^2}_{\text{HUP, G}} = \frac{\hbar^2 \beta}{2} = \frac{\hbar^2\gamma}{4}(2l+1) \, .
\end{equation}

Conversely, under the EUP framework, the momentum operator itself is deformed according to $\hat{p} = -i\hbar(1+\gamma x^2)\frac{d}{dx} - i\hbar \gamma x$. Evaluating its expectation value in the Gaussian state requires computing higher-order spatial moments:
\begin{align}
    \hat{p}^2 &= -\hbar^2\left[(1+\gamma x^2)^2 \frac{d^2}{dx^2} + 4\gamma x(1+\gamma x^2)\frac{d}{dx} + \gamma(1+2\gamma x^2)\right] , \nonumber \\
    \braket{\hat{p}^2}_{\text{G, EUP}} &= -\hbar^2 \left[ \braket{\hat{x}^6}\gamma^2\beta^2 + \braket{\hat{x}^4}(2\gamma\beta^2-5\gamma^2\beta) + \braket{\hat{x}^2}(\beta^2+2\gamma^2-6\beta\gamma) + (\gamma-\beta) \right] \nonumber \\
    &= \frac{\hbar^2\gamma}{2} \frac{2l^2+4l+5}{2l+1} \, .
\end{align}

This expression can be decomposed into the standard HUP contribution plus a subleading geometric correction term:
\begin{equation}
    \braket{\hat{p}^2}_{\text{G, EUP}} = \frac{\hbar^2\gamma}{4}(2l+1) + \frac{\hbar^2\gamma}{4} \frac{2l+9/2}{2l+1} \, .
\end{equation}

In the limit $l \gg 1$ (or $\gamma \to 0$), these two Gaussian momentum variances converge to a common value. This asymptotic agreement extends to the exact EUP ground-state momentum variance derived in Appendix~\eqref{appendix: vargsw}:
\begin{equation}
    \braket{\hat{p}^2}_{\text{EUP}} = \hbar^2\gamma \frac{(l+1)^2}{2l +1} = \frac{\hbar^2\gamma}{4} \frac{(2l+2)^2}{2l +1} = \frac{\hbar^2\gamma}{4}\left[(2l+1) + \frac{1}{2l +1} + 1\right] .
\end{equation}
As $l \gg 1$, all three expressions for the momentum variance asymptotically converge to $\frac{\hbar^2\gamma}{2}l$.

\section{Implications of the Maximum Entropy Principle for EUP-Deformed Quantum Systems}
\label{appendix:MEP_implications}

In this appendix, we investigate the validity and physical implications of the Maximum Entropy Principle (MEP) in Extended Uncertainty Principle (EUP) deformed quantum systems. In Appendix~\eqref{appendix: rcm_rdm_entanglement}, we analyze the Reduced Covariance Matrix (RCM) and Reduced Density Matrix (RDM) formalisms for the Gaussian Reference State (GRS), identifying the parameter regime ($l_- \gg 1$) where both approaches yield identical entanglement entropy and validating the MEP. In Appendix~\eqref{appendix: productstateCHO}, we evaluate the entanglement entropy of an exact EUP product state expanded in a Hermite basis, demonstrating that its entropy is strictly bounded above by the GRS, thereby providing a direct manifestation of the MEP. Finally, in Appendix~\eqref{appendix: rotHerm}, we derive a generalized algebraic identity for products of Hermite polynomials with rotated and scaled arguments, establishing the exact mathematical framework required for tracing out subsystem degrees of freedom.

\subsection{RCM Entanglement vs. RDM Entanglement}
\label{appendix: rcm_rdm_entanglement}

As noted in the main text, the RDM formalism applied to the GRS yields the standard HUP momentum variance because the Gaussian ansatz cannot simultaneously encode the exact EUP momentum variance. In this subsection, we analyze the physical regime where the RDM and RCM approaches yield asymptotically identical results and discuss its implications for the Maximum Entropy Principle.

The normal-mode momentum variances for the GRS are given by:
\begin{equation}
    \braket{\hat{p}_\pm^2}_{\text{G, HUP}} = \frac{\hbar^2\gamma}{4}(2l_\pm+1) \, , \qquad \braket{\hat{p}^2_\pm}_{\text{G, EUP}} = \frac{\hbar^2\gamma}{2} \frac{2l_\pm^2+4l_\pm+5}{2l_\pm+1} \, .
\end{equation}

\begin{figure}[H]
    \centering
    \includegraphics[width=0.32\linewidth]{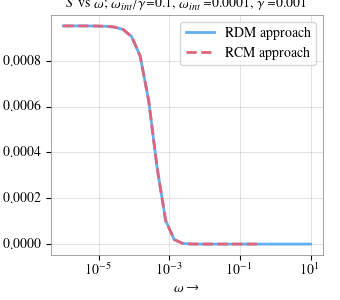}
    \includegraphics[width=0.32\linewidth]{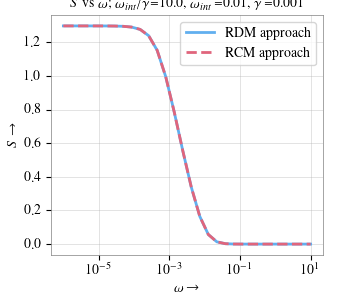}
    \includegraphics[width=0.32\linewidth]{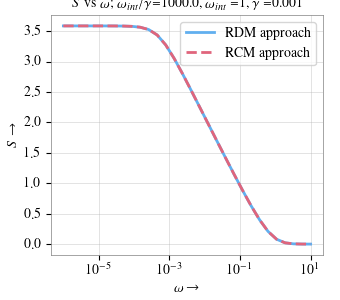}
    \caption{Comparison of RCM and RDM entanglement entropy versus frequency $\omega$, evaluated using $\braket{\hat{p}_1^2}_{\text{G, HUP}}$ as the RCM matrix entry, with $M=\hbar=1$.}
    \label{fig:SvsOmega_RCM_RDM_HUP}
\end{figure}

The element of the RCM corresponding to the momentum variance of the coupled harmonic oscillator is defined as $\braket{\hat{p}_1^2} = \frac{1}{2}\left(\braket{\hat{p}_+^2} + \braket{\hat{p}_-^2}\right)$. In Fig.~\eqref{fig:SvsOmega_RCM_RDM_HUP}, we present numerical plots of the entanglement entropy computed using this matrix entry evaluated under the HUP momentum definition, denoted as $\braket{\hat{p}_1^2}_{\text{G, HUP}}$.

The entropy profiles obtained from the two formalisms match identically. We repeat this calculation using the EUP momentum definition:

\begin{figure}[H]
    \centering
    \includegraphics[width=0.32\linewidth]{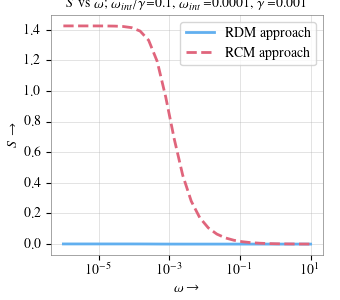}
    \includegraphics[width=0.32\linewidth]{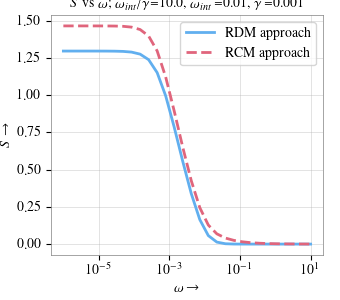}
    \includegraphics[width=0.32\linewidth]{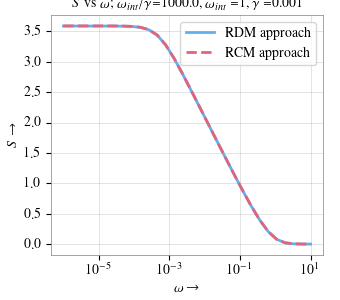}
    \caption{Comparison of RCM and RDM entanglement entropy versus frequency $\omega$, evaluated using $\braket{\hat{p}_1^2}_{\text{G, EUP}}$ as the RCM matrix entry, with $M=\hbar=1$.}
    \label{fig:SvsOmega_RCM_RDM_EUP}
\end{figure}

Since, the RDM depends exclusively on the wave-function representation, the GRS fails to capture the exact EUP momentum variance. A noticeable discrepancy therefore emerges in the regime where the dimensionless parameter $\kappa = \frac{M\omega_{\text{int}}}{\hbar\gamma} \ll 1$. This behavior can be understood analytically. In the IR regime characterized by $\omega = 0$ ($l_+ = 0$), the respective momentum variances reduce to:
\begin{equation}
    \braket{\hat{p}^2_1}_{\text{G, HUP}} = \frac{\hbar^2\gamma}{4}(l_-+1) \, , \qquad \braket{\hat{p}^2_1}_{\text{G, EUP}} = \frac{\hbar^2\gamma}{2} \frac{l_-^2+7l_-+5}{2l_- +1} \, .
\end{equation}

The difference and the ratio between these expressions are:
\begin{align}
    \braket{\hat{p}^2_1}_{\text{G, EUP}} - \braket{\hat{p}^2_1}_{\text{G, HUP}} &= \frac{\hbar^2\gamma}{4} \frac{11l_-+9}{2l_-+1} \, , \\
    \frac{\braket{\hat{p}^2_1}_{\text{G, EUP}}}{\braket{\hat{p}^2_1}_{\text{G, HUP}}} &= \frac{1 + 7/l_- + 5/l_-^2}{\left(1 + 0.5/l_-\right)\left(1 + 1/l_-\right)} \, .
\end{align}

Holding $\omega_{\text{int}}$ fixed and selecting $\gamma \ll 1$ satisfies the condition $l_- \gg 1$. Consequently, the relative difference vanishes asymptotically and the ratio approaches unity. This regime enables a consistent evaluation of the entanglement entropy using both methodologies, establishing a foundation for applying the Maximum Entropy Principle.

The Maximum Entropy Principle requires that the covariance matrix coincides with the exact EUP covariance matrix. While the spatial components are identical between the two matrices, a discrepancy exists in their momentum components. However, evaluating the ratios $\braket{\hat{p}_1^2}_{\text{EUP}}/\braket{\hat{p}^2_1}_{\text{G, EUP}}$ and $\braket{\hat{p}_1^2}_{\text{EUP}}/\braket{\hat{p}^2_1}_{\text{G, HUP}}$ reveals that both quantities converge to unity in the limit $l_- \gg 1$. Consequently, the validity of the Maximum Entropy Principle is preserved.

\subsection{Entanglement Analysis for the Product State}
\label{appendix: productstateCHO}

Although the energy-based limitations of the product state were demonstrated in Appendix~\eqref{appendix:CHO}, its resulting covariance matrix yields exact EUP spatial and momentum variances. Evaluating its entanglement entropy serves as a test of the Maximum Entropy Principle.

Consider the product wave function:
\begin{equation}
\Psi(x_+,x_-) = \frac{N(l_+)N(l_-)}{\left(1+\gamma x_+^2\right)^{1+l_+/2} \left(1+\gamma x_-^2\right)^{1+l_-/2}} \, ,
\end{equation}
where the normalization constant is defined as 
$$
N(l) = \left(\frac{\gamma}{\pi}\right)^{1/4} \left[\frac{\Gamma(l+2)}{\Gamma(l+3/2)}\right]^{1/2} \, .
$$
To define the reduced density matrix, we expand the wave function 
in a Hermite polynomial basis. Setting the parameter $\alpha = \frac{1}{2\braket{\hat{x}^2}_{\text{EUP}}} = \frac{\gamma(2l_++1)}{2}$ connects it to the variance of the EUP-modified $\omega_+$ mode, ensuring maximum overlap with the center-of-mass mode:
\begin{equation}
    \Psi(x_1,x_2) = \sum_{m,n=0}^\infty d_n c_m e^{-\frac{\alpha}{2} (x_1^2 + x_2^2)} H_m(\sqrt{\alpha} x_+ ) H_n(\sqrt{\alpha} x_- ) \, .
\end{equation}
Here, the expansion coefficients are $c_n = \frac{\braket{\phi_n, \psi(l_+)}}{\braket{\phi_n, \phi_n}}$ and $d_n = \frac{\braket{\phi_n, \psi(l_-)}}{\braket{\phi_n, \phi_n}}$. Transforming from normal-mode coordinates to physical coordinates requires the algebraic identity derived in Appendix~\eqref{appendix: rotHerm}:
\begin{equation}
    H_p\left(\frac{x+y}{\sqrt{2}}\right) H_q\left(\frac{x-y}{\sqrt{2}}\right) = \sum_{s=0}^{p+q} Q_{s}^{pq} H_s(x) H_{p+q-s}(y) \, ,
\end{equation}
where,
\begin{equation}
    Q_s^{pq} = \sum_{k=\max(0,s-q)}^{\min(s,p)} \frac{(-1)^{s-k} p! q!}{k! (s-k)! (p-k)! (q+k-s)! \, 2^{(p+q)/2}} \, .
\end{equation}

Tracing out the $x_2$ coordinate yields the reduced density matrix:
\begin{equation}
    \rho(x,x') = \sum_{s=0}^\infty \sum_{b=0}^{\infty} \sum_{r=s}^{\infty} A_{rs} B_{rsb} H_s(\sqrt{\alpha} x) H_{b}(\sqrt{\alpha} x') K_{r-s} \exp\left(-\frac{\alpha(x^2+{x'}^2)}{2}\right) ,
\end{equation}
where $K_{r-s} = (r-s)! \, 2^{r-s} \sqrt{\frac{\pi}{\alpha}}$, and the expansion matrices $A$ and $B$ are defined as:
\begin{align}
    A_{rs} &= \sum_{n=0}^r d_n c_{r-n} Q_{s}^{r-n,n} \, , \\
    B_{rsb} &= \sum_{q=0}^{r+b-s} d_q c_{r+b-q-s} Q_{b}^{r+b-s-q,q} \, .
\end{align}
We express the $n$-th eigenfunction of the reduced density operator in the form:
\begin{equation}
    f_n(x) = \sum_{k=0}^{\infty} D_k^n H_k(\sqrt{\alpha}x) e^{-\alpha x^2/2} \, .
\end{equation}
Solving the integral eigenvalue equation yields:
\begin{equation}
    \sum_{b=0}^\infty J_{kb} \, b! \, 2^b \sqrt{\frac{\pi}{\alpha}} D_b^n = p_n D_k^n \, .
\end{equation}
This represents the matrix eigenvalue equation $\tilde{J}D^n = p_n D^n$, where the eigenvalues $p_n$ correspond to the eigenvalues of $\tilde{J}$, with elements $\tilde{J}_{mn} = J_{mn} \, n! \, 2^n \sqrt{\frac{\pi}{\alpha}}$.

\begin{figure}[H]
    \centering
    \includegraphics[width=0.45\linewidth]{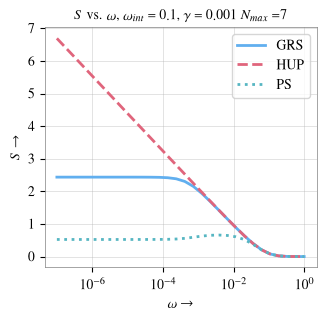}
    \includegraphics[width=0.45\linewidth]{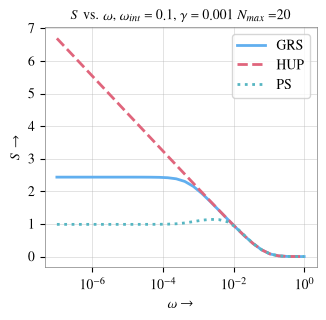}
    \caption{Comparison of entanglement entropy for the product state, Gaussian reference state, and standard HUP framework across different truncation orders.}
    \label{fig:SvswCoupled_diffn}
\end{figure}

The von Neumann entanglement entropy is evaluated as $S = -\sum_{n} p_n \ln{p_n}$. The eigenvalues $p_n$ are determined numerically by diagonalizing $\tilde{J}$ or estimated analytically by truncating the series expansion of $\Psi$ up to a finite order in the Hermite basis $H_n$. Truncating the series at different values of $n$ yields the results shown in Fig.~\eqref{fig:SvswCoupled_diffn}.

The numerical results indicate that the IR divergence reported in Ref.~\cite{chandran_one--one_2020} is suppressed in the EUP framework. Because the product state is constructed from exact EUP eigenstates, it shares the same covariance matrix as the Gaussian Reference State. The observation that the product state's entanglement entropy is strictly smaller than that of the GRS is a physical manifestation of the Maximum Entropy Principle.

\subsection{Derivation of Scaled and Rotated Hermite Polynomial Identities}
\label{appendix: rotHerm}

Starting from the generating function of Hermite polynomials:
\begin{align}
    \sum_{m=0}^{\infty}\sum_{n=0}^{\infty} H_m\left(\frac{x+y}{\sqrt{2}}\right) H_n\left(\frac{x-y}{\sqrt{2}}\right) \frac{t^m s^n}{m!n!} &= \exp\left[ \frac{2(x+y)t}{\sqrt{2}} + \frac{2(x-y)s}{\sqrt{2}} - t^2 - s^2 \right] \nonumber \\
    &= \exp\left[ \frac{2x(t+s)}{\sqrt{2}} - \frac{(t+s)^2}{2} + \frac{2y(t-s)}{\sqrt{2}} - \frac{(t-s)^2}{2} \right] .
\end{align}
The right-hand side can be expressed in terms of $H_p(x)$ and $H_q(y)$:
\begin{equation}
    \sum_{m=0}^{\infty}\sum_{n=0}^{\infty} H_m\left(\frac{x+y}{\sqrt{2}}\right) H_n\left(\frac{x-y}{\sqrt{2}}\right) \frac{t^m s^n}{m!n!} = \sum_{p=0}^\infty\sum_{q=0}^\infty \frac{H_p(x)H_q(y)}{p! q! \, 2^{(p+q)/2}} (t+s)^p (t-s)^q \, .
\end{equation}

Expanding the factors $(t\pm s)$ binomially gives:
\begin{equation}
    \sum_{m=0}^{\infty}\sum_{n=0}^{\infty} H_m\left(\frac{x+y}{\sqrt{2}}\right) H_n\left(\frac{x-y}{\sqrt{2}}\right) \frac{t^m s^n}{m!n!} = \sum_{p=0}^\infty\sum_{q=0}^\infty \frac{H_p(x)H_q(y)}{p! q! \, 2^{(p+q)/2}} \sum_{j=0}^p\sum_{k=0}^q \binom{p}{j}\binom{q}{k}(-1)^{q-k} t^{j+k} s^{p+q-j-k} \, . \label{eqn:Hermpdt1}
\end{equation}
Setting $m = j+k$ and $n = p+q-j-k$, and re-indexing to ensure non-negative indices, we obtain:
\begin{equation}
    \sum_{m=0}^{\infty}\sum_{n=0}^{\infty} H_m\left(\frac{x+y}{\sqrt{2}}\right) H_n\left(\frac{x-y}{\sqrt{2}}\right) \frac{t^m s^n}{m!n!} = \sum_{m=0}^\infty\sum_{n=0}^\infty \sum_{q=0}^{m+n} \sum_{k=\max(0,q-n)}^{\min(q,m)} \frac{(-1)^{q-k} H_{m+n-q}(x) H_q(y) \, t^m s^n}{k! (q-k)! (m-k)! (n+k-q)! \, 2^{(m+n)/2}} \, .
\end{equation}
Equating coefficients of $t^m s^n$ yields:
\begin{align}
    H_m\left(\frac{x+y}{\sqrt{2}}\right) H_n\left(\frac{x-y}{\sqrt{2}}\right) &= \sum_{q=0}^{m+n} \sum_{k=\max(0,q-n)}^{\min(q,m)} \frac{(-1)^{q-k} m! n! \, H_{m+n-q}(x) H_q(y)}{k! (q-k)! (m-k)! (n+k-q)! \, 2^{(m+n)/2}} \nonumber \\
    &= \sum_{q=0}^{m+n} Q_q^{mn} H_{m+n-q}(x) H_q(y) \, .
\end{align}

This identity can be generalized by starting from Eq.~\eqref{eqn:Hermpdt1} and introducing scaling parameters $\alpha$ and $\beta$ into the arguments of the Hermite polynomials:
\begin{align}
    &\sum_{m=0}^{\infty}\sum_{n=0}^{\infty}  H_m\left(\frac{\alpha}{\sqrt{2}}(x+y)\right) H_n\left(\frac{\beta}{\sqrt{2}}(x-y)\right) \frac{t^m s^n}{m!n!} \nonumber \\
    &= \exp\left[s^2 \left(\frac{\beta^2}{\alpha^2} - 1\right)\right] \sum_{m=0}^\infty\sum_{n=0}^\infty \sum_{q=0}^{m+n} \sum_{k=\max(0,q-n)}^{\min(q,m)} (-1)^{q-k} \left(\frac{\beta}{\alpha}\right)^n \frac{H_{m+n-q}(\alpha x) H_q(\alpha y) \, t^m s^n}{k! (q-k)! (m-k)! (n+k-q)! \, 2^{(m+n)/2}} \, .
\end{align}
Expanding the exponential term in powers of $s$ and collecting powers of $s$ and $t$ yields:
\begin{equation}
    H_m\left(\frac{\alpha}{\sqrt{2}}(x+y)\right) H_n\left(\frac{\beta}{\sqrt{2}}(x-y)\right) = \sum_{j=0}^{\lfloor n/2 \rfloor} \sum_{q=0}^{m+n-2j} Q_{mnjq} H_{m+n-2j-q}(\alpha x) H_q(\alpha y) \, ,
\end{equation}
where the generalized expansion coefficients are given by:
\begin{equation}
    Q_{mnjq} = \sum_{l=\max(0,q+2j-n)}^{\min(q,m)} \left(\frac{\beta}{\alpha}\right)^{n-2j} \left(\frac{\beta^2}{\alpha^2} - 1\right)^j \frac{(-1)^{q-l} m! n!}{j! l! (m-l)! (q-l)! (n+l-q-2j)! \, 2^{(m+n-2j)/2}} \, .
\end{equation}
\section{EUP-Deformed 1-D Harmonic chain}
\label{appendix:Hchain}

In this appendix, we analyze the continuum limit of a one-dimensional EUP-deformed harmonic chain to extract the analytical properties of its entanglement entropy in the massless limit ($\omega \to 0$). We explicitly evaluate the continuous mode integrals in terms of complete elliptic integrals and demonstrate how the EUP deformation algebra regulates IR zero-mode divergences. Furthermore, we provide a systematic asymptotic derivation of the von Neumann entropy in the regime of large symplectic eigenvalues ($\nu \gg 1$).

\subsection{Continuum limit of 1-D Harmonic Chain}
\label{appendix: contLimitChain}

Taking the continuum limit of the one-dimensional harmonic chain provides analytical insights into the system's entanglement entropy. In this section, we derive the continuous field limit and obtain an explicit analytical expression for the von Neumann entropy at $\omega = 0$. The normal-mode frequencies of the discrete chain are given by:
\begin{equation}
    \tilde{\omega}_k^2 = \omega^2 + 4\omega_{\text{int}}^2\sin^2\left(\frac{\pi k}{N}\right) , \qquad k \in \{0, 1, \dots, N-1\} \, .
\end{equation}

The exact normal-mode position and momentum variances modified by the EUP framework are:
\begin{equation}
    \braket{\tilde{x}_k^2} = \frac{1}{\gamma (2l_k + 1)} \, , \qquad \braket{\tilde{p}^2_k} = \frac{\hbar^2\gamma}{4}\left[(2l_k+1) + \frac{1}{2l_k + 1} + 1\right] ,
\end{equation}
where $\kappa \equiv M\omega_{\text{int}}/\hbar$, and the parameter $l_k$ is defined via:
\begin{equation}
    2l_k+1 = \sqrt{1 + 4\left(\frac{M\omega}{\hbar \gamma}\right)^2 + \left(\frac{4M\omega_{\text{int}}}{\hbar \gamma}\right)^2 \sin^2\left(\frac{\pi k}{N}\right)} \, .
\end{equation}

In the continuum limit $N \to \infty$, the discrete sum over the $k$-modes transitions into a continuous integral. In the strictly massless limit ($\omega \to 0$), evaluating the symplectic eigenvalue requires the following continuous integrals:
\begin{align}
    \frac{1}{\pi}\int_0^\pi dk \, (2l_k+1) &= \frac{1}{\pi}\int_0^\pi dk \, \sqrt{1 + \left(\frac{4\kappa}{\gamma}\right)^2 \sin^2 k} \nonumber \\
    &= \frac{2\sqrt{1+ 16\kappa^2/\gamma^2}}{\pi} E\left(\frac{1}{\sqrt{1+\gamma^2/(16\kappa^2)}}\right) , \\
    \frac{1}{\pi}\int_0^\pi dk \, \frac{1}{2l_k+1} &= \frac{1}{\pi}\int_0^\pi \frac{dk}{\sqrt{1 + \left(\frac{4\kappa}{\gamma}\right)^2 \sin^2 k}} \nonumber \\
    &= \frac{2}{\pi\sqrt{1+16\kappa^2/\gamma^2}} K\left(\frac{1}{\sqrt{1+\gamma^2/(16\kappa^2)}}\right) ,
\end{align}
where $K$ and $E$ denote the complete elliptic integrals of the first and second kinds, respectively~\cite{Olver:2010ouy}. 

Defining the dimensionless parameter $\eta \equiv \sqrt{1+16\kappa^2/\gamma^2}$, the squared symplectic eigenvalue $\nu^2$ of the continuous field is given by:
\begin{align}
    \nu^2 &= \frac{1}{\pi^2} \int_0^\pi dk \, \braket{\tilde{x}^2(k)} \int_0^\pi dk' \, \braket{\tilde{p}^2(k')} \nonumber \\
    &= \frac{\hbar^2}{4\pi^2} \left[ \frac{2}{\eta} K\left(\frac{1}{\eta}\right) + 4 E\left(\frac{1}{\eta}\right) K\left(\frac{1}{\eta}\right) + \frac{4}{\eta^2} K^2\left(\frac{1}{\eta}\right) \right] .
\end{align}
In the regime where the continuous field approximation applies smoothly ($\kappa/\gamma \gg 1 \implies \eta \gg 1$), the first and third terms become negligible due to their heavily suppressed algebraic prefactors $2/\eta$ and $4/\eta^2$. The primary physical contribution originates from the central cross-term $4 E(1/\eta) K(1/\eta)$. Utilizing $E(0) = \pi/2$, the squared symplectic eigenvalue simplifies to:
\begin{equation}
    \nu^2 \approx \frac{1}{\pi^2} K\left(\frac{1}{\sqrt{1+\gamma^2/(16\kappa^2)}}\right) \approx \frac{1}{\pi^2} \ln\left(\frac{16\kappa}{\gamma}\right) .
    \label{eqn:eigvalMassless2}
\end{equation}
For large symplectic eigenvalues ($\nu \gg 1$), the corresponding von Neumann entanglement entropy follows the asymptotic expansion derived in Appendix~\eqref{appendix: EE_largenu}:
\begin{equation}
    S \approx 1 + \ln\nu \approx 1 - \ln\pi + \frac{1}{2}\ln\left[\ln\left(\frac{16\kappa}{\gamma}\right)\right] .
\end{equation}
Thus, within the EUP framework, the divergent contribution to the entanglement entropy at zero mass is dictated purely by geometry. Standard field-theoretic treatments (such as Ref.~\cite{mallayadiv}) require an ad hoc cutoff function to suppress zero-mode divergences. The double-logarithmic structure $\ln(\ln(16\kappa/\gamma))$ obtained here confirms that the EUP deformation algebra naturally regulates the divergent contributions of the IR zero modes.

\subsection{Entanglement Entropy in the Large-$\nu$ Limit}
\label{appendix: EE_largenu}

The von Neumann entanglement entropy for a single bosonic mode is uniquely determined by its symplectic eigenvalue $\nu$. In physical regimes characterized by strong squeezing or IR zero-mode contributions—such as the continuum massless limit analyzed above—the symplectic eigenvalue becomes large ($\nu \gg 1$). Below, we derive the asymptotic expansion of the entropy function $S(\nu)$ in this limit to justify the relation $S \approx 1 + \ln\nu$ used in Eq.~\eqref{eqn:eigvalMassless2}.

Starting from the exact expression for the von Neumann entropy of a single mode,
\begin{equation}
    S(\nu) = \left(\nu+\frac{1}{2}\right)\ln\left(\nu+\frac{1}{2}\right) - \left(\nu-\frac{1}{2}\right)\ln\left(\nu-\frac{1}{2}\right) ,
\end{equation}
we factor out $\nu$ from the logarithmic arguments:
\begin{equation}
    \ln\left(\nu \pm \frac{1}{2}\right) = \ln\nu + \ln\left(1 \pm \frac{1}{2\nu}\right) .
\end{equation}
Expanding the second term using the Taylor series $\ln(1+x) = x - \frac{x^2}{2} + \mathcal{O}(x^3)$ for $|x| \ll 1$ yields
\begin{equation}
    \ln\left(\nu \pm \frac{1}{2}\right) = \ln\nu \pm \frac{1}{2\nu} - \frac{1}{8\nu^2} + \mathcal{O}(\nu^{-3}) .
\end{equation}
Substituting these expansions into $S(\nu)$ and collecting terms up to order $\mathcal{O}(\nu^{-1})$, we obtain:
\begin{align}
    S(\nu) &= \left(\nu+\frac{1}{2}\right)\left[\ln\nu + \frac{1}{2\nu} - \frac{1}{8\nu^2} + \mathcal{O}(\nu^{-3})\right] - \left(\nu-\frac{1}{2}\right)\left[\ln\nu - \frac{1}{2\nu} - \frac{1}{8\nu^2} + \mathcal{O}(\nu^{-3})\right] \nonumber \\
    &= \ln\nu + 1 + \mathcal{O}\left(\nu^{-2}\right) .
\end{align}
Hence, in the limit $\nu \gg 1$, the von Neumann entropy behaves asymptotically as
\begin{equation}
    S(\nu) \sim 1 + \ln\nu \, .
\end{equation}
\section{EUP-Deformed $(1+1)$-D scalar field theory}
\label{appendix: scalarfield}

Starting from the Klein-Gordon equation \eqref{eqn:KGeqn}, we assume a separable classical field solution of the form $\Phi_n(x,t) = \varphi_n(x) f(t)$, where the spatial mode functions satisfy the eigenvalue equation $\hat{\D}^2 \varphi_n = -\lambda_n^2 \varphi_n$. This completely separates the spatial and temporal degrees of freedom, yielding the general mode expansion:
\begin{equation}
    \Phi(x,t) = \sum_{n=0}^\infty \varphi_n(x) \left( A_n e^{-i\zeta_n t} + B_n e^{i\zeta_n t} \right) ,
\end{equation}
where $\zeta_n^2 = m_f^2 + \lambda_n^2$. Noting that $\varphi_n^*(x)$ also satisfies the spatial eigenvalue equation and imposing the reality condition $\hat{\Phi}^\dagger(x,t) = \hat{\Phi}(x,t)$, the quantum field operator is written as
\begin{equation}
    \hat{\Phi}(x,t) = \sum_{n=0}^\infty \left[ \varphi_n(x) e^{-i\zeta_n t} \hat{A}_n + \varphi_n^*(x) e^{i\zeta_n t} \hat{A}_n^\dagger \right] .
\end{equation}
Taking the time derivative yields the conjugate momentum field operator $\hat{\Pi}(x,t) = \partial_t \hat{\Phi}(x,t)$. Assuming $[\hat{A}_m, \hat{A}_n] = 0$ and $[\hat{A}_m^\dagger, \hat{A}_n^\dagger] = 0$, the equal-time field-momentum commutation relation takes the form
\begin{equation}
    [\hat{\Phi}(x,t), \hat{\Pi}(y,t)] = 2i \sum_{m=0}^\infty \sum_{n=0}^\infty [\hat{A}_m, \hat{A}_n^\dagger] \zeta_m \varphi_n^*(x) \varphi_m(y) e^{-i(\zeta_m - \zeta_n)t} \, .
\end{equation}
Imposing the operator commutator $[\hat{A}_m, \hat{A}_n^\dagger] = \frac{\delta_{mn}}{2\zeta_m}$ reduces the right-hand side to the standard spatial completeness relation:
\begin{equation}
    [\hat{\Phi}(x,t), \hat{\Pi}(y,t)] = i \sum_{n=0}^\infty \varphi_n^*(x) \varphi_n(y) = i \delta(x-y) \, .
\end{equation}
For convenience, we rescale the operator set $\{\hat{A}_n, \hat{A}_n^\dagger\}$ according to
\begin{equation}
    \hat{a}_n = \sqrt{2\zeta_n} \hat{A}_n \, , \qquad \hat{a}_n^\dagger = \sqrt{2\zeta_n} \hat{A}_n^\dagger \, .
\end{equation}
Consequently, the mode-expanded field operator assumes the canonical form
\begin{equation}
    \hat{\Phi}(x,t) = \sum_{n=0}^\infty \frac{1}{\sqrt{2\zeta_n}} \left[ \varphi_n(x) e^{-i\zeta_n t} \hat{a}_n + \varphi_n^*(x) e^{i\zeta_n t} \hat{a}_n^\dagger \right] .
\end{equation}

Alternatively, the continuous mode expansion and creation/annihilation operators can be established using the standard Klein-Gordon scalar product~\cite{Birrell:1982ix,Parker:2009uva,Padmanabhan:2016xjk}. It is easy to see that the resulting commutation relations derived from this scalar product are consistent with the operator quantization used here. Using the definition of the conjugate field, the field Hamiltonian in the EUP framework is expressed as
\begin{equation}
    \hat{H} = \frac{1}{2} \int dx \left[ \hat{\Pi}^2 + (1 + \gamma x^2)^2 (\partial_x \hat{\Phi})^2 + \left[ m_f^2 - \gamma(1 + 2\gamma x^2) \right] \hat{\Phi}^2 \right] .
\end{equation}
To simplify the spatial derivative term, we integrate by parts under the assumption that the field configurations vanish at spatial infinity:
\begin{equation}
    \int_{-\infty}^{\infty} (1 + \gamma x^2)^2 (\partial_x \hat{\Phi})^2 dx = -\int_{-\infty}^{\infty} \hat{\Phi} \frac{d}{dx} \left[ (1 + \gamma x^2)^2 \frac{\partial \hat{\Phi}}{\partial x} \right] dx \, .
\end{equation}
Recalling the definition of the modified spatial differential operator $\hat{\D}^2$, expanding the derivative term yields
\begin{equation}
    \frac{d}{dx} \left[ (1 + \gamma x^2)^2 \frac{\partial \hat{\Phi}}{\partial x} \right] = \hat{\D}^2 \hat{\Phi} + \gamma(1 + 2\gamma x^2) \hat{\Phi} \, .
\end{equation}
Substituting this relation back into the Hamiltonian allows the potential terms proportional to $\gamma(1 + 2\gamma x^2)$ to cancel exactly, reducing the Hamiltonian to
\begin{equation}
    \hat{H} = \frac{1}{2} \int dx \left[ \hat{\Pi}^2 + m_f^2 \hat{\Phi}^2 - \hat{\Phi} \hat{\D}^2 \hat{\Phi} \right] .
\end{equation}

We now substitute the explicit mode expansions for $\hat{\Phi}(x,t)$ and $\hat{\Pi}(x,t)$ in terms of the rescaled ladder operators $\hat{a}_n$ and $\hat{a}_n^\dagger$. Utilizing the spatial eigenvalue equation $\hat{\D}^2 \varphi_n = -\lambda_n^2 \varphi_n$ together with the orthonormality condition $\int \varphi_m^*(x) \varphi_n(x) dx = \delta_{mn}$, the time-dependent cross-terms proportional to $e^{\pm 2i\zeta_n t}$ cancel identically due to the dispersion relation $\zeta_n^2 = \lambda_n^2 + m_f^2$. 

The remaining time-independent terms combine directly to give
\begin{equation}
    \hat{H} = \sum_{n=0}^\infty \frac{\zeta_n}{2} \left( \hat{a}_n^\dagger \hat{a}_n + \hat{a}_n \hat{a}_n^\dagger \right) .
\end{equation}
Applying the canonical commutation relation $[\hat{a}_n, \hat{a}_m^\dagger] = \delta_{nm}$, we obtain the standard normal-ordered form up to an additive zero-point energy contribution:
\begin{equation}
    \hat{H} = \sum_{n=0}^\infty \zeta_n \left( \hat{a}_n^\dagger \hat{a}_n + \frac{1}{2} \right) .
\end{equation}
To verify that the rescaled operators $\hat{a}_n$ and $\hat{a}_n^\dagger$ function as ladder operators that shift the energy eigenvalues of the system, we evaluate their commutation relations with the Hamiltonian $\hat{H}$. Using the normal-ordered Hamiltonian and setting aside the constant zero-point energy term, the commutator $[\hat{a}_n, \hat{H}]$ is
\begin{equation}
    [\hat{a}_n, \hat{H}] = \left[ \hat{a}_n, \sum_{m=0}^\infty \zeta_m \hat{a}_m^\dagger \hat{a}_m \right] = \sum_{m=0}^\infty \zeta_m [\hat{a}_n, \hat{a}_m^\dagger \hat{a}_m] \, .
\end{equation}
Using the operator identity $[\hat{A}, \hat{B}\hat{C}] = [\hat{A}, \hat{B}]\hat{C} + \hat{B}[\hat{A}, \hat{C}]$ along with $[\hat{a}_n, \hat{a}_m^\dagger] = \delta_{nm}$, the commutator evaluates to
\begin{equation}
    [\hat{a}_n, \hat{a}_m^\dagger \hat{a}_m] = [\hat{a}_n, \hat{a}_m^\dagger] \hat{a}_m + \hat{a}_m^\dagger [\hat{a}_n, \hat{a}_m] = \delta_{nm} \hat{a}_m \, .
\end{equation}
Substituting this result back into the summation gives
\begin{equation}
    [\hat{a}_n, \hat{H}] = \zeta_n \hat{a}_n \, .
\end{equation}
Taking the Hermitian conjugate of this relation yields the corresponding commutator for the creation operator:
\begin{equation}
    [\hat{a}_n^\dagger, \hat{H}] = -\zeta_n \hat{a}_n^\dagger \, .
\end{equation}
These two algebraic relations establish the ladder behavior of the operators. Specifically, if $|\mathcal{E}\rangle$ is an energy eigenstate satisfying $\hat{H}|\mathcal{E}\rangle = \mathcal{E}|\mathcal{E}\rangle$, acting with the commutator gives
\begin{equation}
    \hat{H}(\hat{a}_n|\mathcal{E}\rangle) = (\hat{a}_n\hat{H} - \zeta_n\hat{a}_n)|\mathcal{E}\rangle = (\mathcal{E} - \zeta_n)(\hat{a}_n|\mathcal{E}\rangle) \, .
\end{equation}
Thus, $\hat{a}_n$ lowers the total energy of the state by $\zeta_n$. Similarly, for the creation operator,
\begin{equation}
    \hat{H}(\hat{a}_n^\dagger|\mathcal{E}\rangle) = (\mathcal{E} + \zeta_n)(\hat{a}_n^\dagger|\mathcal{E}\rangle) \, ,
\end{equation}
confirming that $\hat{a}_n$ and $\hat{a}_n^\dagger$ act as energy lowering and raising operators, respectively.

\end{document}